# A Unified Stochastic Model of Handover Measurement in Mobile Networks

Van Minh Nguyen, *Member, IEEE*, Chung Shue Chen, *Member, IEEE*, and Laurent Thomas

*Abstract*—Handover measurement is responsible for finding a handover target and directly decides the performance of mobility management. It is governed by a complex combination of parameters dealing with multicell scenarios and system dynamics. A network design has to offer an appropriate handover measurement procedure in such a multiconstraint problem. This paper proposes a unified framework for the network analysis and optimization. The exposition focuses on the stochastic modeling and addresses its key probabilistic events, namely: 1) suitable handover target found; 2) service failure; 3) handover measurement triggering; and 4) handover measurement withdrawal. We derive their closed-form expressions and provide a generalized setup for the analysis of handover measurement failure and target cell quality by the best signal quality and level crossing properties. Finally, we show its application and effectiveness in today's 3GPP-LTE cellular networks.

## I. INTRODUCTION

IN MOBILE cellular networks, a user may travel across different cells during a service. Handover (HO) that switches the user's connection from one cell to another is an essential function. Technology advancement is expected to minimize service interruption and to provide seamless mobility management [1]. A handover procedure contains two functions, namely handover *measurement* and handover *decision-execution*. The measurement function is responsible for monitoring the service quality from the serving cell and finding a suitable neighboring cell for handover. Handover decision-execution is made after the measurement function: It decides whether or not to execute a handover to the neighboring cell targeted by the measurement function, and then coordinates multiparty handshaking among the user and cell sites to have HO execution fast and transparent. In mobile-assisted network-controlled

V. M. Nguyen is with Advanced Studies, R&D Department, Sequans Communications, 92073 Paris-La Défense, France (e-mail: vanminh.nguyen@sequans.com).

C. S. Chen is with the Network Technologies Department, Alcatel-Lucent Bell Labs, 91620 Nozay, France, and also with the Laboratory of Information, Networking and Communication Sciences (LINCS), 75013 Paris, France (e-mail: cs.chen@alcatel-lucent.com).

L. Thomas is with the Network Technologies Department, Alcatel-Lucent Bell Labs, 91620 Nozay, France (e-mail: laurent.thomas@alcatel-lucent.com).

handover [2], which is recommended by all cellular standards for its operational scalability and effectiveness, the mobile is in charge of the HO measurement function. It measures the signal quality of neighboring cells and reports the measurement result to the network to make an HO decision.

It is clear that the quality of the handover target cell is directly determined by handover measurement function. Moreover, the handover measurement is performed during the active state of the mobile in the network, called *connected-mode*, and would impact the ongoing services. Advanced wireless broadband systems such as 3G and 4G [3] allow adjacent cells operating in a common frequency band, and thereby enable the measurement of several neighboring cells simultaneously. This results in enhanced handover measurement. Its efficiency is primarily determined by the number of cells that the mobile is able to measure simultaneously during a measurement period, which is called the mobile's *measurement capability*. For instance, a 3GPP-LTE (Long Term Evolution) compliant terminal is required be able to measure eight cells in each measurement period of 200 ms [4].

Handover is essentially an important topic for mobile networks and has received many investigations. However, most prior arts were concentrated on the handover control problem and its decision algorithms; see, e.g., [5] and [6]. The handover measurement function has received much less attention [7], and most investigations and analysis are through simulations over some case studies or selected scenarios. It is difficult to design a few representative simulation scenes from which one can draw conclusion that is applicable universally to various system settings. To reduce the dependence on simulation that is often heavy and very inefficient for large networks, here we derive closed-form expressions for handover measurement via stochastic geometry and level-crossing analysis techniques.

While a handover control problem can be studied conventionally in a simplified model of two cells—see, e.g., [8] and [9]—in which a handover decision is made by assigning the mobile to one of them, the handover measurement problem involves a much more complex system in which the signal quality of the best cell among a large number of cells needs to be determined with respect to the experienced interference and noise. This often incurs modeling and analysis difficulty, especially when stochastic parameters are introduced to better describe wireless channels and network dynamics. Moreover, cellular standards have introduced many parameters to control the handover measurement operation, e.g., 3GPP specifies more than 10 measurement reporting events for 3G networks and also many for LTE. The complexity makes handover measurement analysis in general difficult. There lacks a clear model and explicit framework of handover measurement, which is essential for network design and analytical optimization.

In this paper, we study the handover measurement of a generic mobile cellular network with an arbitrary number of base stations. For the physical reality and mathematical convenience, we use the popular Poisson point process model for the locations of the base stations [10] and derive closed-form expressions for the handover measurement including the best signal quality, failure probability, target cell quality, etc. As an application of the above results, we can analyze the performance of handover measurement in LTE and for example investigate how optimal today's design is or could be.

To the best of our knowledge, the work presented in this paper is the first that provides a thorough stochastic analysis of handover measurement. The main contributions of the paper are as follows.

- We establish a unified framework of handover measurement in multicell systems with exact details and modelling for the analytical design and optimization of practical networks.
- We derive the handover measurement state diagram and determine their closed-form expressions to facilitate system analysis and performance evaluation by stochastic geometry and level crossing analysis.
- We apply the above results and investigate the handover measurement in today's LTE with respect to mobile's measurement capability and standard system configuration, and finally provide a set of universal curves that one can use to tradeoff parameters of design preference.

The rest of the paper is organized as follows. Section II gives a review of related work of the topic. Section III describes the HO measurement procedure and system model. Section IV presents some basic definitions of the HO events. Section V explains the resulting state diagrams in detail. Section VI derives their closed-form expressions. Section VII applies the proposed framework to study the HO measurement in LTE networks and presents its numerical results. Finally, Section VIII draws the conclusion.

## II. RELATED WORK

The state of art and research challenges of handover management in mobile WiMAX networks for 4G are discussed in [11]. Adaptive channel scanning is proposed in [12] such that scanning intervals are allocated among mobile stations with respect to the QoS requirements of supported applications and to trade off user throughput and fairness. The efficiency of scanning process in handover procedure is studied in [13]. Results show that for a minimal handover interruption time, the mobile station should perform association with the neighbor base station that provides the best signal quality.

In [14], a comparative study of WCDMA handover measurement procedure on its two measurement reporting options is conducted. Simulation results show that periodic reporting outperforms event-triggered mode, but at the expense of increased signaling cost. For LTE systems, the impact of time-to-trigger, user speed, handover margin, and measurement bandwidth to handover measurement are investigated through simulations in [15]–[17], respectively. Moreover, handover measurement with linear and decibel signal averaging is studied in [18]. Simulation shows that both of them have very similar result. Handover measurement for soft handover is addressed in [19]. The authors propose clustering method using network self-organizing map [20] combined with data mining.

Note that most existing results investigated handover measurement procedure through simulations. Although each simulation could study the impact of a specific setting or parameter to the system performance, there is a lack of a unified analytical framework with tractable closed-form expressions on this topic. This paper establishes a complete stochastic model and mathematical characterization of the handover measurement with explicit formulation of the involved probabilistic events. It is a generalization of the study in [7], which only contains a basic setup of major handover events and is thus limited to continual handover measurement such as intrafrequency handover measurement in WCDMA systems [21]. Moreover, it investigates the influence of different factors on the handover measurement for network optimization.

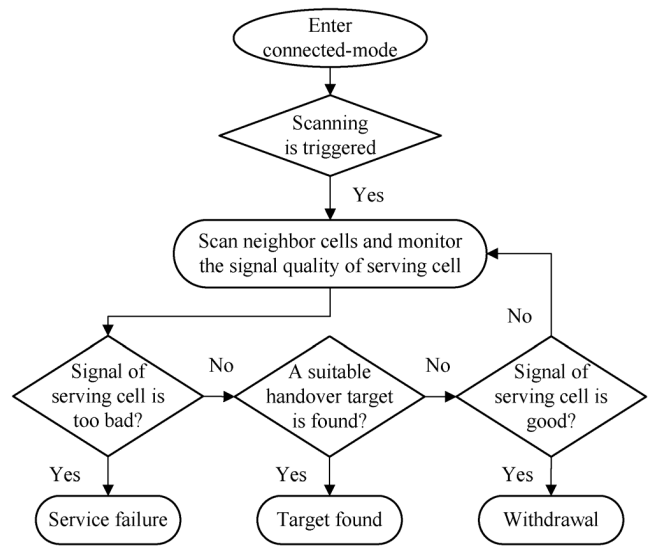

Fig. 1. Handover measurement procedure in mobile networks.

## III. SYSTEM DESCRIPTION

To begin with, we explain handover measurement procedure. Some technical definitions and mathematical notations are necessary and defined here.

### A. Handover Measurement Procedure

The handover measurement in mobile networks, also called *scanning*, can be described in Fig. 1. The mobile station, also known as user equipment (UE), starts scanning neighboring cells as soon as a predefined condition is triggered, e.g., when its received pilot power drops below a certain threshold. Note that the UE needs a certain time duration for measuring the signal quality of a neighboring cell. This time duration is called *measurement period*, denoted by $T_{\text{meas}}$. During each measurement period $[(m-1)T_{\text{meas}}, mT_{\text{meas}}]$, where $m = 1, 2, \ldots$ and time instant 0 refers to the moment when the mobile enters into connected-mode (one may refer to RRC Connected-mode in LTE [22] or the state when the mobile has an active radio connection with the serving base station), the mobile would measure the signal quality of a number of $k$ neighboring cells, where $k$ is known as the mobile's measurement capability. By signal quality, we mean the *signal-to-interference-plus-noise ratio* (SINR) of the received signal, which is an important metric for coverage, capacity, and throughput. By the measurement, a UE obtains the signal quality of neighboring cells by the end of each time period, denoted by $m \times T_{\text{meas}}$. For notational simplicity, in case of no ambiguity, we will use $m$ and $mT_{\text{meas}}$ interchangeably and use $[m-1, m]$ to refer to the measurement period $[(m-1)T_{\text{meas}}, mT_{\text{meas}}]$ accordingly.

During the measurement of neighboring cells, by the nature of wireless link, the serving cell's signal quality may undergo fluctuations that may lead to various possible consequences, e.g., call drop or service failure, a decision of switching to a better neighboring cell, withdraw the scanning, etc. It is worth noting that the signal quality of the serving cell affects the user's quality of service (QoS) in a timescale as short as a symbol time, which is usually much shorter than $T_{\text{meas}}$. During each measurement period $[m-1, m]$, if the signal quality of the serving cell is too bad, the scanning would end in *failure* as a call drop or service interrupt occurs. In such a case, the mobile will then perform standardized procedure of service recovery, called radio link reestablishment in 3GPP-LTE. Otherwise, the scanning ends in *success* if a suitable handover target was identified. By contrast, the mobile may *withdraw* the scanning to reduce the scanning overheads if the signal quality of the serving cell becomes good enough. If that is not the case, the mobile will *continue* the scanning and keep monitoring the signal quality received from the serving cell. This is the exact handover measurement procedure.

### B. Wireless Link Model

The underlying network is composed of a number of base stations (or say transmitting nodes) on a two-dimensional Euclidean plane $\mathbb{R}^2$. We consider that the transmitting nodes are spatially distributed according to a Poisson point process of intensity $\lambda$ for physical reality and mathematical convenience [10].

Considering a nominated user located at $\mathbf{y} \in \mathbb{R}^2$, the signal power received from a base station (BS) $i$ located at $\mathbf{x}_i \in \mathbb{R}^2$ is expressible as

$$P_i(\mathbf{y}) = \frac{P_{\text{tx}} Z_i}{l(\|\mathbf{y} - \mathbf{x}_i\|)} \tag{1}$$

where $0 < P_{\text{tx}} < \infty$ is the base station's transmit power, $l(\cdot)$ is the path loss between the BS and UE, which is defined by typical power-law model expressible as

$$l(d) = (\max(d, d_{\text{min}}))^\beta, \qquad \text{for } d \in \mathbb{R}_+$$

where $d_{\text{min}}$ is some nonnegative constant, and $\beta > 2$ is the path-loss exponent, and the random variables $\{Z_i, i = 1, 2, \ldots\}$ account for fading effects, which could be fast fading, shadowing, or both. However, since fast fading usually varies much faster than the handover delay supported by mobile network standards, in this study $\{Z_i\}$ refer to lognormal shadowing that is expressible as

$$Z_i = 10^{\frac{X_i}{10}} \tag{2}$$

where the random variables $\{X_i, i = 1, 2, \ldots\}$ are independently and identically distributed (i.i.d.) according to a Gaussian distribution with zero mean and standard deviation $0 < \sigma_X < \infty$.

The signal quality of base station $i$ expressed in terms of signal-to-interference-plus-noise ratio is given by

$$Q_i(\mathbf{y}) = \frac{P_i(\mathbf{y})}{(N_0 + I_i(\mathbf{y}))} \tag{3}$$

### TABLE I
### Basic Notations

| Symbol | : Definition |
|---|---|
| $P_i, Q_i$ | : signal power, signal quality of transmitting node $i$ |
| $l(\cdot), d_{\text{min}}, \beta$ | : pathloss function, excluding distance, pathloss exponent |
| $Z_i, X_i$ | : shadowing in linear scale, in decibel scale (dB) |
| $R_X$ | : autocorrelation function of $X_i(t)$ |
| $D_\xi, U_\xi$ | : down excursion, up excursion of process $\xi(t)$ |
| $T_{\text{meas}}, m$ | : measurement period, measurement instant |
| $k$ | : measurement capacity (cells per measurement period) |
| $\gamma_{\text{req}}$ | : required minimum level of HO target |
| $\gamma_{\text{min}}, \tau_{\text{min}}$ | : min. level and min. duration of to service failure |
| $\gamma_t, \tau_t$ | : triggering level, triggering duration |
| $\gamma_{\text{w}}, \tau_{\text{w}}$ | : withdrawal threshold and the experienced duration |
| $\delta_{\text{scan}}, \delta_{\text{HO}}$ | : scan margin, and handover margin |
| $findtarget_m$ | : finding a HO target at measurement instant $m$ |
| $fail_m$ | : service failure in period $[m-1, m]$ |
| $trig_m$ | : scanning trigger in period $[m-1, m]$ |
| $wdraw_m$ | : scanning withdrawal in period $[m-1, m]$ |
| $\pi_m(i, j)$ | : transition probability from state $i$ at instant $m-1$ |
| | : to state $j$ at instant $m$ |
| $\pi_m$ | : distribution of states probability at instant $m$ |
| $\mathbf{M}_m$ | : transition matrix of handover measurement process |
| $\mathbf{N}_m$ | : transition matrix of mobile in connected-mode |
| $\mathcal{F}$ | : probability of failed handover measurement |
| $\mathcal{S}$ | : probability of successful handover measurement |
| $\mathcal{Q}$ | : expected signal quality of the target cell |

where $N_0$ is the thermal noise average power and $I_i(\mathbf{y}) \triangleq \sum_{j \neq i} P_j(\mathbf{y})$ is the sum of interference. For notational simplicity, we let $P_i(\mathbf{y}) \triangleq P_i(\mathbf{y})/N_0$ with a little abuse of notation, and thus rewrite (3) as

$$Q_i(\mathbf{y}) = \frac{P_i(\mathbf{y})}{(1 + I_i(\mathbf{y}))}. \tag{4}$$

In the temporal domain, we also consider that $X_i(t)$ is stationary with auto-correlation function $R_X(\tau)$. Furthermore, for a time-varying process $\xi(t)$, we denote

$D_\xi(\gamma, \tau, [t_a, t_b])$ :   event that $\xi(t) < \gamma$ for $t \in [t_0, t_0 + \delta_t]$
  with $\delta_t \geq \tau$ and $t_0 + \tau \in [t_a, t_b]$.

Similarly, we define

$U_\xi(\gamma, \tau, [t_a, t_b])$ :   event that $\xi(t) > \gamma$ for $t \in [t_0, t_0 + \delta_t]$
  with $\delta_t \geq \tau$ and $t_0 + \tau \in [t_a, t_b]$.

Note that in the above definitions, the starting time $t_0$ of the crossing events does not necessarily belong to the time window $[t_a, t_b]$. Without loss of generality, the inequality signs $<$ and $>$ can be simply replaced by $\leq$ and $\geq$, respectively. We refer the reader to [23] for more details on the level-crossing properties of a stationary process.

### IV. Basic Definitions

Following the above notations, we provide the mathematical definitions of the handover measurement events. Table I summarizes our notations for brevity.

### A. Suitable Handover Target Found

A *suitable* handover target is a candidate neighboring cell to which the serving base station would consider to handover the mobile. A handover is then conducted by a handover *execution* procedure.

A suitable handover target needs to satisfy some *necessary* conditions. One necessary condition is that the signal quality of the suitable handover target must be greater than a required threshold $\gamma_{\text{req}}$. A handover decision process may consider more criteria to refine the selection depending on the control algorithm used by the base station, for example considering the current load of the candidate handover target and/or the relative signal quality between the candidate handover target and the serving cell. Notice that here we deal with the handover measurement function whose role is to *find a suitable handover target* and prevent service failure; criteria for handover decision process are thus not of our interest here.

Since in each measurement period a mobile scans $k$ cells and the cell with best signal quality is preferable, the event of having a suitable handover target found at time instant $m$ can be defined by

$$\text{findtarget}_m(k) \triangleq \{Y_k \geq \gamma_{\text{req}}\} \qquad (5)$$

where

$$Y_k \triangleq \max_{i=1,\cdots,k} Q_i \qquad (6)$$

refers to the best signal quality received from the $k$ cells scanned.

### B. Service Failure

In wireless communications, the signal may undergo time-varying fading and other impairments like interference such that its instantaneous signal quality fluctuates. This would result in packet errors when the signal quality is poor. Techniques such as interleaving, automatic repeat request (ARQ), and hybrid-ARQ (H-ARQ) are often used to maintain the communication reliability. These techniques are however effective to recover data only when the packet error rate is relatively low. When the SINR stays below a minimum allowable level, say $\gamma_{\text{min}}$, for long time such that successive bursts are erroneous, those error-fighting techniques do not help any more, leading to a service failure. For instance, LTE considers that a radio link failure is to be detected if a maximum number of retransmissions (under ARQ or H-ARQ mechanism) is reached. Therefore, it is more appropriate and also generic to incorporate a minimum duration $\tau_{\text{min}}$ when defining the event. A service failure during $[m-1, m]$ is thus defined by an excursion of the serving cell's signal quality falling below the minimum tolerable level $\gamma_{\text{min}}$ with minimum-duration $\tau_{\text{min}}$

$$\text{fail}_m \triangleq D_{Q_0}\left(\gamma_{\text{min}}, \tau_{\text{min}}, [m-1, m]\right) \qquad (7)$$

where $Q_0(t)$ denotes the SINR received from the serving cell at time $t$. Fig. 2 gives an illustration. A service failure occurs at instant "D," where the serving cell's signal quality drops below $\gamma_{\text{min}}$ for a duration $\tau_{\text{min}}$. Note that when $\tau_{\text{min}} = 0$, the definition in (7) corresponds to an instantaneous SINR outage, which is a special case of our expression.

### C. Scanning Trigger

Since handover measurement introduces overheads such as gaps in data transmission or mobile's resource consumption, one expects to perform a handover measurement only when the signal quality of the serving cell is really bad. Since an SINR may cross and stay below or above a threshold instantaneously or only for a very short duration, the handover measurement should be triggered only if the serving cell's signal quality drops below a certain threshold, say $\gamma_t$, for a certain period, denoted by $\tau_t$. It is clear that if these two parameters are not appropriately configured, it may happen that a service failure occurs before the handover measurement initiation. In such a case, the mobile has to conduct a link reestablishment procedure, which is not favorable. One can see that the handover measurement is triggered during period $[m-1, m]$ if the serving cell's signal quality is worse than threshold $\gamma_t$ during at least $\tau_t$ and when no service failure occurs in this period, i.e.,

$$\text{trig}_m \triangleq D_{Q_0}\left(\gamma_t, \tau_t, [m-1, m]\right) \wedge \neg\text{fail}_m \qquad (8)$$

where $\wedge$ and $\neg$ stand for logical `and` and logical `negation`, respectively. It is clear that $\gamma_t$ should be set greater than $\gamma_{\text{min}}$. For illustration, in Fig. 2, the handover measurement is triggered at instant "A" and also at instant "C," respectively.

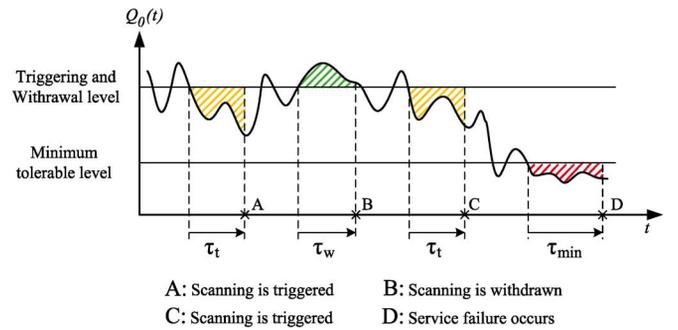

Fig. 2. Level crossing events experienced by a mobile user.

### D. Scanning Withdrawal

Similarly, the handover measurement should be withdrawn when the signal quality of the serving cell becomes good. Precisely, it should be withdrawn if the serving cell's signal quality is higher than a threshold $\gamma_w$ for a certain period $\tau_w$. Thus, the event of scanning withdrawal during period $[m-1, m]$ is expressible as

$$\text{wdraw}_m \triangleq U_{Q_0}\left(\gamma_w, \tau_w, [m-1, m]\right). \qquad (9)$$

In Fig. 2, we consider that $\gamma_w = \gamma_t$ and $\tau_w = \tau_t$. Since $Q_0(t)$ crosses over $\gamma_t$ and stays above over a duration $\tau_t$, the scanning process is canceled at instant "B."

### V. HANDOVER STATE DIAGRAM

In consequence, a handover measurement would result in *failure* if a service failure occurs—see, e.g., instant "D" in Fig. 2—and *success* if a suitable handover target is found before its occurrence. It is particularly of primary importance to determine the probability of *handover measurement failure* and also related metrics.

The mobile's handover measurement activities during connected-mode in a cellular network can be described by the four states capsuled in Fig. 3. States Scan and NoScan describe whether the mobile is scanning neighboring cells or not. States Fail and CellSwitch describe if the mobile is encountering a service failure or if it is being switched to another cell, respectively. For ease of analytical development, we number the four states as 1–4, as shown in Fig. 3. The transition probability from state $i$ at instant $m-1$ to state $j$ at instant $m$, where $i, j \in \{1, 2, 3, 4\}$, is denoted by $\pi_m(i, j)$.

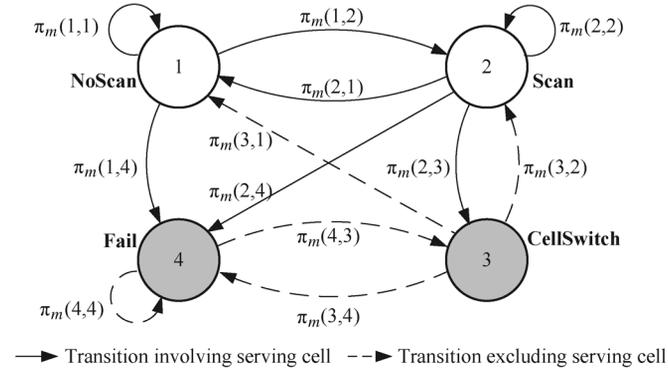

Fig. 3. State diagram of the mobile in connected-mode in the network.

Denote by

$$\pi_m := (\pi_m(1), \pi_m(2), \pi_m(3), \pi_m(4))$$

the row vector of the state probability at instant $m$. Vector $\pi_0$ corresponds to the starting instant 0. In the following, we explain the details of the state diagram one by one, which is the core for all the performance evaluation.

### A. State Analysis: `NoScan`

From state `NoScan`, a mobile will start a handover measurement and enter state `Scan` if the triggering condition occurs. It will fall into state `Fail` if the mobile encounters a service failure during the period $[m-1, m]$. Otherwise, it remains in state `NoScan`.

Note that a mobile does not scan neighboring cells when being in state `NoScan`. As a consequence, there is no transition from `NoScan` to `CellSwitch`, unless the network may force the mobile to connect to another cell out of the procedure. One can see that reducing scanning overhead by increasing the state probability of `NoScan` for example by raising the triggering threshold $\gamma_{\mathrm{t}}$ and/or prolonging the triggering duration $\tau_{\mathrm{t}}$ will increase the risk of service failure.

The transition probabilities $\pi_m(1, j)$, for $j = 1, \dots, 4$, are thus expressible as

$$
\begin{aligned}
\pi_m(1, 2) &= \mathbb{P}(\mathrm{trig}_m) \\
\pi_m(1, 3) &= 0 \\
\pi_m(1, 4) &= \mathbb{P}(\mathrm{fail}_m) \\
\pi_m(1, 1) &= 1 - \pi_m(1, 2) - \pi_m(1, 4).
\end{aligned}
\tag{10}
$$

### B. State Analysis: `Scan`

In state `Scan`, a mobile performs handover measurement as shown in Fig. 2, while the received signal may undergo level crossings. Following Fig. 1, the transition probabilities from state `Scan` to the other can be written as

$$\pi_m(2, 4) = \mathbb{P}(\mathrm{fail}_m)$$

$$
\begin{aligned}
\pi_m(2, 3) &= (1 - \pi_m(2, 4)) \, \mathbb{P}(Y_k \geq \gamma_{\mathrm{req}}) \\
\pi_m(2, 1) &= (1 - \pi_m(2, 4)) \, \mathbb{P}(Y_k < \gamma_{\mathrm{req}}) \mathbb{P}(\mathrm{wdraw}_m) \\
\pi_m(2, 2) &= 1 - (\pi_m(2, 1) + \pi_m(2, 3) + \pi_m(2, 4))
\end{aligned}
\tag{11}
$$

where $\mathbb{P}(Y_k \geq \gamma_{\mathrm{req}})$ is the probability of finding a suitable handover target with $Y_k$ denoted by (5).

### C. State Analysis: `CellSwitch`

In state `CellSwitch`, a mobile is switched to the identified target cell. If the signal quality of the new serving cell is too bad or if the handover execution procedure cannot be completed, the mobile will encounter a service failure. In such a case, the mobile falls into state `Fail`. On the other hand, given the signal quality of the new serving cell, when triggering condition holds, the mobile will then go into a `Scan` state and start to scan neighboring cells again; otherwise, it will go into state `NoScan`. Thus, the transition probabilities from state `CellSwitch` are expressible as

$$
\begin{aligned}
\pi_m(3, 2) &= \mathbb{P}\left(\mathrm{trig}_m^*\right) \\
\pi_m(3, 4) &= \mathbb{P}\left(\mathrm{fail}_m^*\right) \\
\pi_m(3, 1) &= 1 - \pi_m(3, 2) - \pi_m(3, 4)
\end{aligned}
\tag{12}
$$

where the events $\mathrm{trig}_m^*$ and $\mathrm{fail}_m^*$ refer to scanning triggering and service failure, respectively, corresponding to the signal quality received from the new serving cell, denoted by $Q_0^*$, where the superscript "*" is used to refer to a new serving cell.

### D. State Analysis: `Fail`

In state `Fail`, a mobile will reinitiate a network admission procedure or conduct link reestablishment to recover the ongoing service from the interruption. The mobile scans possible neighboring cells; when a suitable cell is found, the mobile will go into state `CellSwitch` so as to connect to the identified cell. The signal quality of the suitable cell is required to be greater than or equal to the minimum tolerable level $\gamma_{\min}$. Otherwise, the mobile keeps scanning to find a suitable cell, during which the service is in failure status. As a result, we have

$$
\begin{aligned}
\pi_m(4, 3) &= \mathbb{P}(Y_k \geq \gamma_{\min}) \\
\pi_m(4, 4) &= 1 - \pi_m(4, 3).
\end{aligned}
\tag{13}
$$

### E. State Transition Matrix

By the result of (10)–(13), the state diagram of a *mobile user* in the network, as illustrated in Fig. 3, can be represented by its state transition matrix expressible as

$$
\mathbf{N}_m \triangleq \begin{pmatrix}
\pi_m(1, 1) & \pi_m(1, 2) & 0 & \pi_m(1, 4) \\
\pi_m(2, 1) & \pi_m(2, 2) & \pi_m(2, 3) & \pi_m(2, 4) \\
\pi_m(3, 1) & \pi_m(3, 2) & 0 & \pi_m(3, 4) \\
0 & 0 & \pi_m(4, 3) & \pi_m(4, 4)
\end{pmatrix}.
\tag{14}
$$

Notice that we will determine each $\pi_m(i, j)$ and derive their closed-form expressions explicitly in Section VI.

To represent the time evolution of the state transitions, Let

$$\mathbf{N}^{(m)} \triangleq \mathbf{N}_1 \times \cdots \times \mathbf{N}_m$$

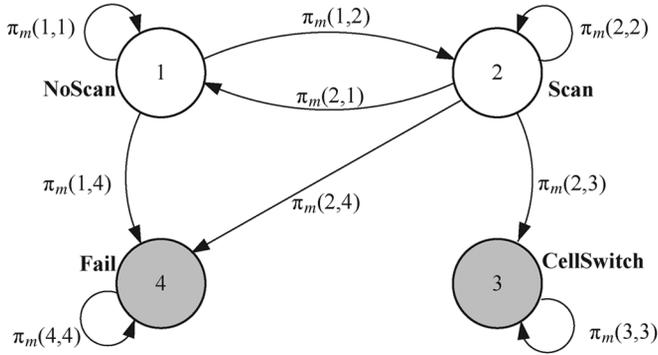

Fig. 4. State diagram of the mobile in handover measurement.

where $\times$ is the inner matrix product. It is clear that $\mathbf{N}^{(m)}(i, j)$ is the transition probability from state $i$ at instant 0 to state $j$ at instant $m$.

Our objective is to derive the state diagram of the *handover measurement*. Note that a mobile is connected to a current serving cell when being in one of two states: `NoScan` or `Scan`. In contrast, the mobile enters `CellSwitch` or `Fail`. One can see that $\pi_m(1, j)$ and $\pi_m(2, j)$ depend on the signal quality of the current serving cell, whereas $\pi_m(3, j)$ and $\pi_m(4, j)$ do not since they occur after the connection with the serving cell was released in case of cell switching or was interrupted in case of service failure. In the latter case, the mobile may proceed with a radio link reestablishment to resume service with a cell. From the viewpoint of mobility management, this cell is considered as a new cell even if it may be the last serving cell prior to the service interruption. However, it is important to distinguish between these two types of states for mathematical derivation. As shown in Fig. 3, `CellSwitch` and `Fail` are shadowed, and their state transitions are drawn in dashed lines. On the other hand, we have the state transitions from `NoScan` and `Scan` drawn in solid lines.

Note that the mobile only performs the handover measurement function when it is in state `Scan`. States `CellSwitch` and `Fail` are outcomes. From the view of a serving cell, the state diagram in Fig. 3 should be refined as Fig. 4, in which the dashed-line transitions are excluded and both `CellSwitch` and `Fail` are absorbing states. Fig. 4 corresponds to the state diagram of the mobile in a handover measurement procedure. The resulting transition matrix is thus given by

$$\mathbf{M}_m \triangleq \begin{pmatrix} \pi_m(1,1) & \pi_m(1,2) & 0 & \pi_m(1,4) \\ \pi_m(2,1) & \pi_m(2,2) & \pi_m(2,3) & \pi_m(2,4) \\ 0 & 0 & 1 & 0 \\ 0 & 0 & 0 & 1 \end{pmatrix}. \quad (15)$$

Let

$$\mathbf{M}^{(m)} \triangleq \mathbf{M}_1 \times \cdots \times \mathbf{M}_m.$$

$\mathbf{M}^{(m)}(i, j)$ is the transition probability of the handover measurement from state $i$ at instant 0 to state $j$ at instant $m$.

The state probability distribution $\pi_m$ at any instant $m$ is thus given by

$$\pi_m = \pi_0 \times \mathbf{M}^{(m)}. \quad (16)$$

Since the starting instant 0 corresponds to the moment when the mobile enters into connected-mode, the initial state probability distribution $\pi_0$ is given as

$$\pi_0 = (1 - \pi_0(2), \pi_0(2), 0, 0) \quad (17)$$

where

$$\pi_0(2) = \mathbb{P}(\mathrm{trig}_0) \quad (18)$$

is the probability that the handover measurement is triggered at instant 0. The above formulation allows evaluating various quantities of interest, including key performance metrics of handover measurement in Section V-F.

### F. Performance Metrics of HO Measurement

Let $t_c$ be the time interval during which the mobile has an active connection with its serving cell, and $m_c = \lceil t_c/T_{\mathrm{meas}} \rceil$, where $\lceil x \rceil$ is the smallest integer greater than or equal to $x$. Notice that $m_c$ is the corresponding number of measurement periods.

As aforementioned, a handover measurement would result in two outcomes that are *failure* if a service failure occur during the handover measurement, and *success* if a suitable handover target can be found in time. The probability of *handover measurement failure*, denoted by $\mathcal{F}$, is of key concern. With the notation developed above, we have

$$\mathcal{F} = \pi_{m_c}(4). \quad (19)$$

Similarly, the probability of *handover measurement success*, denoted by $\mathcal{S}$, is given by

$$\mathcal{S} = \pi_{m_c}(3). \quad (20)$$

The above expressions take into account all the involved factors including the terminal's measurement capability $k$, system specified measurement time $T_{\mathrm{meas}}$, scan triggering and withdrawal parameters $(\gamma_{\mathrm{t}}, \tau_{\mathrm{t}})$ and $(\gamma_{\mathrm{w}}, \tau_{\mathrm{w}})$, as well as HO target level $\gamma_{\mathrm{req}}$ and service failure thresholds $(\gamma_{\mathrm{min}}, \tau_{\mathrm{min}})$.

Note that the time interval $t_c$ can be treated as a deterministic constant or as a random variable. In the latter case, let $\Lambda$ be its distribution. The above metrics can be rewritten as

$$\mathcal{F} = \int_0^\infty \pi_{\lceil \frac{t}{T_{\mathrm{meas}}} \rceil}(4) \Lambda(\mathrm{d}t) \quad (21)$$

and

$$\mathcal{S} = \int_0^\infty \pi_{\lceil \frac{t}{T_{\mathrm{meas}}} \rceil}(3) \Lambda(\mathrm{d}t) \quad (22)$$

respectively. In the literature, $\Lambda$ has been modeled by some known distributions such as a truncated log-logistic distribution. The interested reader is referred to [24] for further information.

Intuitively, $\mathcal{F}$ represents the probability that a service failure occurs before a suitable cell is identified, whereas $\mathcal{S}$ indicates the probability that the system goes into the handover decision-execution phase. It is desirable to have $\mathcal{F}$ as small as possible. To do so, one may consider simply having low handover target level $\gamma_{\mathrm{req}}$. However, this may result in handover to cells of low signal quality. It is thus important to assess the performance of the handover measurement by the *target cell quality*, which is expressible as

$$\mathcal{Q} \triangleq \mathcal{S} \times \mathbb{E}\{Y_k | Y_k \geq \gamma_{\mathrm{req}}\} \tag{23}$$

where

$$\mathbb{E}\{Y_k | Y_k \geq \gamma_{\mathrm{req}}\} = \gamma_{\mathrm{req}} + \int\limits_{\gamma_{\mathrm{req}}}^{\infty} \frac{\overline{F}_{Y_k}(y)\mathrm{d}y}{\overline{F}_{Y_k}(\gamma_{\mathrm{req}})}$$

with tail distribution $\overline{F}_{Y_k}$ of $Y_k$. Note that a suitable target cell is given by the best cell among $k$ cells scanned and provided that its signal quality is greater than or equal to $\gamma_{\mathrm{req}}$.

## VI. ANALYTICAL CLOSED-FORM EXPRESSIONS

By the results of Sections IV and V, we can derive the probabilities of $\mathrm{findtarget}_m$, $\mathrm{fail}_m$, $\mathrm{trig}_m$, and $\mathrm{wdraw}_m$, cf. (6)–(9), respectively. First, we derive $\mathrm{findtarget}_m$, i.e., $\mathbb{P}(Y_k \geq \gamma)$, for any threshold $\gamma > 0$. Then, we derive $\mathbb{P}(\mathrm{fail}_m)$ and $\mathbb{P}(\mathrm{trig}_m)$ built on the down-crossing events $\mathrm{fail}_m$ and $\mathrm{trig}_m$, respectively. Finally, by the up-crossing event $\mathrm{wdraw}_m$, we determine $\mathbb{P}(\mathrm{wdraw}_m)$ to complete the analytical formulation.

### A. Probability of Finding a Suitable Cell

To determine the probability $\mathbb{P}(Y_k \geq \gamma)$, one needs to define the set of candidate cells from which $k$ cells would be taken for the handover measurement. By today's cellular standards [2], [21], [25], there are two cases:
- *limited* candidate set;
- *unlimited* candidate set.

In the former, a mobile only scans neighboring cells of a predefined set that contains a limited number of potential candidates, say $N_{\mathrm{cell}}$ cells. The set in practice corresponds to the neighbor cell list (NCL) as used in GSM, WCDMA, and WiMAX with $N_{\mathrm{cell}} = 32$. In the case of unlimited candidate set, the mobile is allowed to scan any cell in the network. However, since a network may have a very large number of cells, scanning without restriction would introduce unaffordable overheads. Therefore, new broadband cellular systems use a set of, say $N_{\mathrm{CSID}}$, cell synchronization identities (CSIDs), to label cells from this finite set. Since this set of $N_{\mathrm{CSID}}$ CSIDs are shared among all the cells, two cells having the same CSID must be spatially separated far enough so as to avoid any confusion. When required to scan $k$ cells, a mobile just picks $k$ out of the total $N_{\mathrm{CSID}}$ CSIDs without a predefined NCL, and then conducts standardized cell synchronization and measurement. An example using this mechanism is LTE that defines 504 physical cell identifiers (PCIs) that serve as CSIDs. The mobile performs the cell measurement autonomously without the need of a preconfigured cell set such as the NCL used in predecessor systems, for the generality.

We determine $\mathbb{P}(Y_k \geq \gamma)$ in both cases and complete the details in the following.

*1) Case of Unlimited Candidate Set:* Each neighboring cell is scanned with equal probability $\rho_k = k/N_{\mathrm{CSID}}$. This set of the scanned cells is in other words a thinning of $\mathbb{R}^2$ with retention probability $\rho_k$. Notice that this set of the scanned cells, say $S_k$, may have more than $k$ cells, and this efficiently describes the real situation where the mobile may detect several cells which have the same CSID. In practice, an LTE eNodeB relies on an automatic neighbor relation table to map a reported PCI to a unique cell global identifier (CGI). Whenever a PCI conflict is detected where two cells having the same PCI are found, the eNodeB may request the UE to perform an explicit CGI acquisition of the PCI in question, which may take a long time, and updates its neighbor relation table.

In consequence, we rewrite (6) as

$$Y_k \equiv \max_{i \in S_k} Q_i \tag{24}$$

and by definition

$$\mathbb{P}(Y_k > \gamma) = \overline{F}_{Y_k}(\gamma) \tag{25}$$

where $\overline{F}_{Y_k}(\cdot)$ is the tail distribution function of $Y_k$. Applying [26, Corollary 5] for the tail distribution of $Y_k$, we have the following proposition.

*Proposition 1:* With the system model and notation as described above, consider the case of unlimited candidate set with $N_{\mathrm{CSID}}$ cell synchronization identities. Assume $l(d) = d^\beta$, then $\mathbb{P}(Y_k > 0) = 1$, and for $\gamma > 0$

$$\mathbb{P}(Y_k > \gamma)$$
$$= \int\limits_{\gamma}^{\infty}\int\limits_{0}^{\infty} \frac{e^{-C_1 w^\alpha - C_2(w,u)}}{\pi}$$
$$\times \left[ -\frac{1+\gamma}{\gamma} \times \cos\left(C_1 w^\alpha \tan\left(\frac{\pi\alpha}{2}\right) + C_3(w,u) + C_4(w,u)\right) \right.$$
$$\left. + \cos\left(C_1 w^\alpha \tan\left(\frac{\pi\alpha}{2}\right) + C_3(w,u) - wu\right) \right] \mathrm{d}w\mathrm{d}u \tag{26}$$

where $\alpha = 2/\beta$, $\rho_k = k/N_{\mathrm{CSID}}$, $C_1 = (1 - \rho_k)\delta$, and

$$C_2(w,u) = \rho_k c_\alpha \frac{{}_1F_2\left(-\frac{\alpha}{2}; \frac{1}{2}, 1 - \frac{\alpha}{2}; -\frac{u^2 w^2}{4}\right)}{u^\alpha}$$
$$C_3(w,u) = \rho_k c_\alpha \frac{\alpha w}{1-\alpha} \frac{{}_1F_2\left(\frac{1-\alpha}{2}; \frac{3}{2}, \frac{3-\alpha}{2}; -\frac{u^2 w^2}{4}\right)}{u^{\alpha-1}}$$
$$C_4(w,u) = w\left(1 - u(1+\gamma)/\gamma\right)$$

in which ${}_1F_2$ denotes the hypergeometric function, and

$$\delta = c_\alpha \Gamma(1-\alpha)\cos\left(\frac{\pi\alpha}{2}\right) \tag{27}$$

with $\Gamma(\cdot)$ denoting the gamma function, and

$$c_\alpha = \pi\lambda P_{\mathrm{tx}}^\alpha \exp\left(\frac{\alpha^2 \sigma_Z^2}{2}\right)$$

with $\sigma_Z = \sigma_X(\log 10/10)$.

*Proof:* By the discussion of (24), $Y_k$ is the best signal quality of a thinning of $\mathbb{R}^2$ with retention probability $\rho_k = k/N_{\mathrm{CSID}}$. Under assumptions that $P_{\mathrm{tx}}$ is a finite positive constant and $Z_i = 10^{X_i/10}$ with $X_i$ being Gaussian and $0 < \sigma_X < \infty$, $P_{\mathrm{tx}}Z_i$ admits a continuous density and $0 < \mathbb{E}\{(P_{\mathrm{tx}}Z_i)^\alpha\} < \infty$. Provided $l(d) = d^\beta$, $\mathbb{P}(Y_k > \gamma)$ is then given by [26, Corollary 5]. ∎

*2) Case of Limited Candidate Set:* Consider $\widehat{B}$ a disk-shaped network area with radius

$$R_{\widehat{B}} = \sqrt{\frac{N_{\mathrm{cell}}}{(\pi\lambda)}}. \tag{28}$$

Under the assumption that base stations are distributed according to a Poisson point process, $\widehat{B}$ has on average $N_{\text{cell}}$ base stations. Thus, by approximating the region of the $N_{\text{cell}}$ neighboring cells by $\widehat{B}$, we have the tail distribution $\mathbb{P}(Y_k > \gamma)$ directly given by [26, Theorem 3]. In particular, we can have some more tractable expression of $\mathbb{P}(Y_k > \gamma)$ by considering the following two cases: scattered networks like rural macro cellular networks where intersite distance is large such that the network density $\lambda$ is small, and dense networks like urban small cell networks where a large number of cells are deployed to support dense traffic such that $\lambda$ is large.

For small $\lambda$, we can have $R_{\widehat{B}} \approx \infty$, i.e., $\widehat{B}$ can be approximated by $\mathbb{R}^2$. Similarly, let $S_k$ be a thinning on $\widehat{B}$ with retention probability

$$\rho_k = \frac{k}{N_{\text{cell}}} \tag{29}$$

such that $S_k$ has on average $k$ cells. The probability of finding a target cell can be obtained according to Proposition 1.

For large $\lambda$ (e.g., a dense network), the approximation $R_{\widehat{B}} \approx \infty$ may be not applicable. In addition, under the assumption that base stations are distributed according to a Poisson point process, the probability that a base station is found very close to a given user may be significant. The unbounded path-loss model (i.e., $d_{\min} = 0$) may be no longer suitable because the effect of its singularity is now nonnegligible. Therefore, bounded path loss with $d_{\min} > 0$ is considered. The probability of finding a suitable cell is given by the following proposition.

*Proposition 2:* With system model and notation as described above, consider the case of *limited* candidate set with $N_{\text{cell}}$, and large $\lambda$. Let $\widehat{B}$ be a disk-shaped network area of radius $R_{\widehat{B}} = \sqrt{N_{\text{cell}}/(\pi\lambda)}$. Then, the following applies.

(i) $Y_k$ is the best signal quality received from $\hat{k}$ cells uniformly selected from $\widehat{B}$ where $\hat{k} = \min(k, N_{\text{cell}})$.

(ii) Assume $d_{\min} > 0$, for $\gamma \geq 0$

$$\mathbb{P}(Y_k > \gamma) \approx \int_\gamma^\infty \left\{ g(u) \int_0^\infty \frac{2}{\pi w} e^{-\delta w^\alpha} \sin\left(w\frac{u - \gamma}{2\gamma}\right) \right.$$
$$\left. \times \cos\left(wu + w\frac{u - \gamma}{2\gamma} - \delta w^\alpha \tan\frac{\pi\alpha}{2}\right) dw \right\} du \tag{30}$$

where approximation holds under the condition that $P_{\text{tx}} d_{\min}^{-\beta}$ is large, $\delta$ is given by (27), and

$$g(x) = \hat{k} \cdot f_P(x) \cdot F_P^{\hat{k}-1}(x), \qquad \text{for } \hat{k} \geq 1$$

with

$$F_P(x) = c\left(\frac{K_1(x)}{a^\alpha} - \frac{K_2(x)}{b^\alpha} - \frac{e^w K_3(x)}{x^\alpha} + \frac{e^w K_4(x)}{x^\alpha}\right)$$

where $w = 2\sigma_X^2/\beta^2$, $a = P_{\text{tx}} R_{\widehat{B}}^{-\beta}$, $b = P_{\text{tx}} d_{\min}^{-\beta}$, $c = P_{\text{tx}}^\alpha (R_{\widehat{B}}^2 - d_{\min}^2)^{-1}$, and $K_j$, $j = 1, \ldots, 4$, refer to the lognormal distributions of parameters $(\mu_j, \sigma_Z)$ with

$$\mu_1 = \log a \qquad \mu_3 = \mu_1 + 2\sigma_Z^2/\beta$$
$$\mu_2 = \log b \qquad \mu_4 = \mu_2 + 2\sigma_Z^2/\beta$$

and $f_P(x) = dF_P(x)/dx$.

*Proof:* Assertion (i) follows from the above discussion considering that the mobile station scans at most $N_{\text{cell}}$ by its measurement capacity $k$. Under the assumption $d_{\min} > 0$, (30) is given following [27, Theorem 2]. ∎

In Proposition 2, $P_{\text{tx}} d_{\min}^{-\beta}$ is nothing but the average signal power received at the excluding distance $d_{\min}$. The average here is with respect to the unit mean fading $Z_i$. The approximation condition that $P_{\text{tx}} d_{\min}^{-\beta}$ is large implies that the excluding distance should be small compared to the cell size.

### B. Probability of Service Failure: $\mathbb{P}(\text{fail}_m)$

Recall (7), where $Q_0(t)$ is the SINR received from serving cell at time $t$, $Q_0(t) < \gamma_{\min}$ is equivalent to $X(t) < \hat{\gamma}_{\min}(t)$ in a logarithmic representation, where

$$\hat{\gamma}_{\min}(t) \triangleq 10 \log_{10}\left(\gamma_{\min} \frac{l(d(t))}{P_{\text{tx}}}(1 + I(t))\right) \tag{31}$$

by substituting (1) and (2) into (4). Note that the excursions of nonstationary process $Q_0(t)$ below threshold $\gamma_{\min}$ are now represented by the excursions of a stationary normal process $X(t)$ [cf. (2)] below the time-varying level $\hat{\gamma}_{\min}(t)$. This transformation would greatly facilitate the coming derivation.

Let $T_{\min}$ be the length of an excursion of $X(t)$ below $\hat{\gamma}_{\min}(t)$. Following (7), $\text{fail}_m$ is thus expressible as an excursion of $X(t)$ below threshold $\hat{\gamma}_{\min}(t)$ with $T_{\min}$ longer than $\tau_{\min}$, i.e.,

$$\text{fail}_m = \{X(t) \text{ stays below } \hat{\gamma}_{\min}(t) \text{ during } T_{\min} \geq \tau_{\min},$$
$$\text{for } t \in [(m-1)T_{\text{meas}} - \tau_{\min}, mT_{\text{meas}}]\} \tag{32}$$

where the considered time window is $[(m-1)T_{\text{meas}} - \tau_{\min}, mT_{\text{meas}}]$ as the failure will occur at a instant $t_0 + \tau_{\min}$ anterior to $(m-1)T_{\text{meas}}$ if the excursion starts at $t_0 < (m-1)T_{\text{meas}} - \tau_{\min}$.

We will use the level crossing properties of $X(t)$ to derive the probability of event $\text{fail}_m$. Given a constant level $\gamma$, we can have the following results.

*Lemma 3 ([23, p. 194]):* Write $C_\gamma$ the number of crossing of $X(t)$ of level $\gamma$ during a unit time

$$\mathbb{E}C_\gamma = \frac{1}{\pi}\sqrt{\frac{\lambda_2}{\lambda_0}} \exp\left(-\frac{\gamma^2}{2\lambda_0}\right) \tag{33}$$

with

$$\lambda_0 = R_X(0) \quad \text{and} \quad \lambda_2 = -R_X''(\tau)|_{\tau=0}.$$

$\mathbb{E}C_\gamma < +\infty$ if and only if $\lambda_2 < +\infty$. ∎

In addition, let $U_\gamma$ and $D_\gamma$ be the number of up-crossings and the number of down-crossings of $X(t)$ of level $\gamma$ during a unit time, respectively. One can find that [23, p. 197]

$$\mathbb{E}U_\gamma = \mathbb{E}D_\gamma = \frac{\mathbb{E}C_\gamma}{2}. \tag{34}$$

*Proposition 4:* With $X(t)$ described above, for constants $\gamma$ and $\tau$, define

$$V(\gamma, \tau) \triangleq \mathbb{P}(X(t) \text{ stays above } \gamma \text{ during at least } \tau).$$

Consider the following assumptions on $R_X(\tau)$.

(i) There exists finite $\lambda_2$, and $a > 1$

$$R_X(\tau) = 1 - \frac{\lambda_2\tau^2}{2} + O\left(\tau^2 |\log|\tau||^{-a}\right), \qquad \text{as } \tau \to 0. \tag{35}$$

(ii) There exist finite $\lambda_2$ and finite $\lambda_4$

$$R_X(\tau) = 1 - \frac{\lambda_2}{2!}\tau^2 + \frac{\lambda_4}{4!}\tau^4 + o(\tau^4), \qquad \text{as } \tau \to 0. \tag{36}$$

(iii) There exists $a > 0$

$$R_X(\tau) = O(\tau^{-a}), \qquad \text{as} \tau \to +\infty. \qquad (37)$$

Then, the following applies.

- For $\gamma \to +\infty$, under conditions (35) and (37)

$$V(\gamma, \tau) = \mathbb{E}U_\gamma \cdot \left( \tau \exp(-A_\gamma \tau^2) + \sqrt{\frac{\pi}{A_\gamma}} Q(\sqrt{2A_\gamma}\tau) \right)$$

where

$$A_\gamma = \frac{\pi}{4} \left( \frac{\mathbb{E}U_\gamma}{Q\left(\frac{\gamma}{\sigma_X}\right)} \right)^2 \quad \text{and} \quad Q(x) \triangleq \int_x^\infty \frac{e^{-t^2/2}}{\sqrt{2\pi}} dt.$$

- For $\gamma \to -\infty$, under conditions (36) and (37)

$$V(\gamma, \tau) = \mathbb{E}U_\gamma \cdot \exp(-\mu\tau) \cdot (\tau + 1/\mu)$$

where $\mu = \mathbb{E}U_\gamma / Q(\gamma/\sigma_X)$.

*Proof:* Please see Appendix B. ∎

Notice that under condition (37), the wide-sense stationary process $X(t)$ is mean-ergodic [28]. Thus, for all intervals $[t_1, t_2]$, one can have

$$\mathbb{P}(X(t) \text{ stays above } \gamma \text{ during at least } \tau, t \in [t_1, t_2])$$
$$= \mathbb{P}(X(t) \text{ stays above } \gamma \text{ during at least } \tau).$$

The result of Proposition 4 is thus applicable for the probability of excursions with minimum required duration considering a finite time interval.

Using Proposition 4, we can obtain $\mathbb{P}(\text{fail}_m)$ for a constant $\hat{\gamma}_{\min}$. However, note that $\hat{\gamma}_{\min}(t)$ is time-varying due to $I(t)$ and $d(t)$. Under the model described in Section III, interference $I$ can be modeled as a shot noise on $\mathbb{R}^2$ [29]. For $l(d) = d^\beta$, its characteristic function is expressible as [26], [30]

$$\phi_I(w) = \exp\left(-\delta|w|^\alpha \left[1 - j\text{sign}(w)\tan\left(\frac{\pi\alpha}{2}\right)\right]\right) \qquad (38)$$

where $\delta$ is given by (27). By the assumption that $\beta > 2$, $0 < \alpha < 1$ since $\alpha = 2/\beta$. In consequence, one can see that $\phi_I(w)$ is absolutely integrable. By the inverse formula [31, Theorem 3], the probability density function (pdf) of $I$ can be obtained

$$f_I(x) = \frac{1}{\pi} \int_0^\infty e^{-\delta w^\alpha} \cos\left(\delta \tan\left(\frac{\pi\alpha}{2}\right) w^\alpha - xw\right) dw. \qquad (39)$$

On the other hand, $\tau_{\min}$ and $T_{\text{meas}}$ are typically about a few hundreds of milliseconds (e.g., $T_{\text{meas}} = 200$ ms under LTE standard [22]). The interval $T_{\text{meas}} + \tau_{\min}$ is so short that wherein the distance between the mobile and its serving base station can be considered constant. By the above results, we can have the following proposition.

*Proposition 5:* With the system described above, assume that $R_X(\tau)$ satisfies conditions (35)–(37), and $l(d) = d^\beta$. Then

$$\mathbb{P}(\text{fail}_m \mid d(t) = d_m) = \int_0^\infty V(-\hat{\gamma}_{\min}, \tau_{\min}) f_I(x) dx \qquad (40)$$

where $f_I$ is given by (39), $V$ is given by Proposition 4, and

$$\hat{\gamma}_{\min} \triangleq 10 \log_{10}\left(\frac{\gamma_{\min} d_m^\beta}{P_{\text{tx}}}(1+x)\right). \qquad (41)$$

Moreover, if $d(t) \approx d_m$ for $t \in [(m-1)T_{\text{meas}} - \tau_{\min}, mT_{\text{meas}}]$, one can have

$$\mathbb{P}(\text{fail}_m) \approx \mathbb{P}(\text{fail}_m \mid d(t) = d_m), \qquad \text{accordingly.}$$

*Proof:* Under the considered assumptions, $I$ admits density $f_I$. Hence, we write

$$\mathbb{P}(\text{fail}_m \mid d(t) = d_m)$$
$$= \int_0^\infty \mathbb{P}(\text{fail}_m \mid I(t) = x, d(t) = d_m) f_I(x) dx$$

in which by (32)

$$\mathbb{P}(\text{fail}_m \mid I(t) = x, d(t) = d_m)$$
$$= \mathbb{P}(X(t) \text{ stays below } \hat{\gamma}_{\min} \text{ during } T_{\min} \geq \tau_{\min})$$
$$\overset{(*)}{=} \mathbb{P}(X(t) \text{ stays above } -\hat{\gamma}_{\min} \text{ during } T_{\min} \geq \tau_{\min})$$

where $(*)$ is by the fact that normal process $X(t)$ is statistically symmetric around its zero mean. The result thus follows using Proposition 4. ∎

*Remark 6:* If one may consider users moving at very high speeds such that $d(t)$ changes significantly during the time interval $T_{\text{meas}} + \tau_{\min}$, $\mathbb{P}(\text{fail}_m)$ can then be computed by a knowledge of the distribution of $d(t)$ with a corresponding mobility model. Here, we make the above approximation for simplification.

### C. Probability of Scanning Trigger: $\mathbb{P}(\text{trig}_m)$

By the definition of the scanning trigger event $\text{trig}_m$ given in (8), we have

$$\mathbb{P}(\text{trig}_m) = \mathbb{P}(D_{Q_0}(\gamma_t, \tau_t, [m-1, m]) \wedge \neg\text{fail}_m)$$
$$= \mathbb{P}(D_{Q_0}(\gamma_t, \tau_t, [m-1, m]))$$
$$\quad - \mathbb{P}(D_{Q_0}(\gamma_t, \tau_t, [m-1, m]) \wedge \text{fail}_m). \qquad (42)$$

Similar to the treatment on $D_{Q_0}(\gamma_{\min}, \tau_{\min}, [m-1, m])$ in Section VI-B for computing $\mathbb{P}(\text{fail}_m)$, and the treatment on $D_{Q_0}(\gamma_t, \tau_t, [m-1, m])$, by the result of Proposition 5, the first term on the right-hand side of (42) is given by

$$\mathbb{P}(D_{Q_0}(\gamma_t, \tau_t, [m-1, m]) \mid I(t) = x, d(t) = d_m) = V(-\hat{\gamma}_t, \tau_t) \qquad (43)$$

where

$$\hat{\gamma}_t \triangleq 10 \log_{10}\left(\frac{\gamma_t d_m^\beta}{P_{\text{tx}}}(1+x)\right). \qquad (44)$$

For the second term on the right-hand side of (42), we have

$$\mathbb{P}(D_{Q_0}(\gamma_t, \tau_t, [m-1, m]) \wedge \text{fail}_m \mid I(t) = x, d(t) = d_m)$$
$$= \mathbb{P}(X(t) \text{ stays below } \hat{\gamma}_{\min} \text{ during } T_{\min} \geq \tau_{\min},$$
$$\quad \text{and } X(t) \text{ stays below } \hat{\gamma}_t \text{ during } T_t \geq \tau_t). \qquad (45)$$

This turns out to be the probability of successive excursions of two adjacent levels. The following result is useful.

*Lemma 7 ([32]):* With $X(t)$ described above, for $\gamma_2 \geq \gamma_1$, $\tau_1 \geq 0$, and $\tau_2 \geq 0$, let $A := A_{\gamma_1}$

$$\tau_1^* := \frac{\gamma_1(\gamma_2 - \gamma_1)}{A}$$

and

$$\tau_2^* = \sqrt{\tau_1^2 - \tau_1^*}, \qquad \text{for } \tau_1^2 \geq \tau_1^*.$$

Define

$W(\gamma_1, \tau_1, \gamma_2, \tau_2) := \mathbb{P}\left(X(t) \text{ stays above } \gamma_1 \text{ during at least}\right.$

$\left. \tau_1 \text{ and } X(t) \text{ stays above } \gamma_2 \text{ during at least } \tau_2\right).$

Put $W := W(\gamma_1, \tau_1, \gamma_2, \tau_2)$ for simplicity. If $R_X(\tau)$ satisfies conditions (35)–(37), then we have the following as $\gamma_1 \to +\infty$.

- For $\tau_1^2 \leq \tau_1^*$ and $\tau_2 = 0$

$$W = \mathbb{E}U_{\gamma_1} \frac{e^{A\tau_1^2}}{e^{2A\tau_1^*}} \sqrt{\frac{\pi}{4A}}.$$

- For $\tau_1^2 \leq \tau_1^*$ and $\tau_2 > 0$

$$W = \mathbb{E}U_{\gamma_1} \frac{e^{A\tau_1^2}}{e^{2A\tau_1^*}} \left( \frac{\tau_2}{e^{A\tau_2^2}} + \sqrt{\frac{\pi}{4A}} \operatorname{erfc}(\sqrt{A}\tau_2) \right).$$

- For $\tau_1^2 > \tau_1^*$ and $\tau_2 = 0$

$$W = \frac{\mathbb{E}U_{\gamma_1}}{e^{A\tau_1^2}} \sqrt{\frac{\pi}{4A}}.$$

- For $\tau_1^2 > \tau_1^*$ and $0 < \tau_2 \leq \tau_2^*$

$$W = \frac{\mathbb{E}U_{\gamma_1}}{e^{A\tau_1^2}} \left( \tau_2^* + \sqrt{\frac{\pi}{4A}} \frac{\operatorname{erfc}(\sqrt{A}\tau_2^*)}{\exp\left(-A(\tau_2^*)^2\right)} \right).$$

- For $\tau_1^2 > \tau_1^*$ and $\tau_2 > \tau_2^*$

$$W = \frac{\mathbb{E}U_{\gamma_1}}{e^{A\tau_1^2}} \left( \frac{\tau_2}{e^{A\left(\tau_2^2 - (\tau_2^*)^2\right)}} + \sqrt{\frac{\pi}{4A}} \frac{\operatorname{erfc}(\sqrt{A}\tau_2)}{e^{-A(\tau_2^*)^2}} \right). \quad \blacksquare$$

By the above analysis, we have the following conclusion.

*Proposition 8:* Assume that $R_X(\tau)$ satisfies conditions (35)–(37), and $l(d) = d^\beta$. Considering that $\gamma_t \geq \gamma_{\min}$, we have

$\mathbb{P}\left(\text{trig}_m \mid d(t) = d_m\right)$

$$= \int_0^\infty \left( V(-\hat{\gamma}_t, \tau_t) - W(-\hat{\gamma}_t, \tau_t, -\hat{\gamma}_{\min}, \tau_{\min}) \right) f_I(x) \mathrm{d}x$$

where $f_I$, $V$, and $W$ are given by (39), Proposition 4, and Lemma 7, respectively, with $\hat{\gamma}_{\min}$ and $\hat{\gamma}_t$ given by (41) and (44), respectively.

*Proof:* Following the above assumptions, $I$ has pdf $f_I$ as given by (39). Hence, we have

$\mathbb{P}\left(\text{trig}_m \mid d(t) = d_m\right)$

$$= \int_0^\infty \mathbb{P}\left(\text{trig}_m \mid I(t) = x, d(t) = d_m\right) f_I(x) \mathrm{d}x.$$

Regarding (45), as $X(t)$ is statistically symmetric around its zero mean, one can have

$\mathbb{P}\left(D_{Q_0}\left(\gamma_t, \tau_t, [m-1, m]\right) \wedge \text{fail}_m \mid I(t) = x, d(t) = d_m\right)$

$= \mathbb{P}\left(X(t) \text{ stays above } -\hat{\gamma}_{\min} \text{ during} T_{\min} \geq \tau_{\min}, \right.$

$\left. \text{and } X(t) \text{ stays above } -\hat{\gamma}_t \text{ during } T_t \geq \tau_t\right)$

where noting that $-\hat{\gamma}_t \leq -\hat{\gamma}_{\min}$, considering $\gamma_t \geq \gamma_{\min}$. This is obtainable by Lemma 7. Using this and (43), the result follows.

## TABLE II
### Deployment Scenarios

| Parameter | | Assumption |
|---|---|---|
| Urban macro-cell | Path loss ($d$ in km) | $l(d) = 128.1 + 37.6 \log_{10} d$ |
| | Transmit power | $P_{BS} = 43$ dBm |
| | Antenna pattern | Omnidirectional |
| | Cell radius | $R = 1000$ m |
| | User's velocity | $v = 50$ km/h |
| Rural macro-cell | Path loss ($d$ in km) | $l(d) = 95.5 + 34.1 \log_{10} d$ |
| | Transmit power | $P_{BS} = 46$ dBm |
| | Antenna pattern | Omnidirectional |
| | Cell radius | $R = 1732$ m |
| | User's velocity | $v = 130$ km/h |
| Shadowing | Standard deviation | $\sigma_X = 10$ dB |
| | Decorr. distance | $d_c = 50$ m |
| Noise | Noise density | $= -174$ dBm/Hz |
| | UE noise figure | $N_F = 9$ dB |

### D. Probability of Scanning Withdrawal: $\mathbb{P}(\text{wdraw}_m)$

The probability of $\text{wdraw}_m$ defined in (9) can be obtained by the same technique used in Section VI-B. Note that $Q_0(t) \geq \gamma_w$ is equivalent to $X(t) \geq \hat{\gamma}_w(t)$ with

$$\hat{\gamma}_w(t) \triangleq 10 \log_{10} \left( \gamma_w \frac{l\left(d(t)\right)}{P_{tx}} \left(1 + I(t)\right) \right).$$

The scanning withdrawal $\text{wdraw}_m$ is then expressible as an up-excursion of $X(t)$ above the level $\hat{\gamma}_w(t)$ for $T_w \geq \tau_w$

$\text{wdraw}_m = \{X(t) \text{ stays above } \hat{w}_{\min}(t) \text{ during } T_w \geq \tau_w,$

$\left. \text{for } t \in [(m-1)T_{\text{meas}} - \tau_w, mT_{\text{meas}}]\right\}.$

Thus, similar to Proposition 5, we can have Proposition 9.

*Proposition 9:* With the system described above, assume that $R_X(\tau)$ satisfies conditions (35)–(37), and $l(d) = d^\beta$. Then

$$\mathbb{P}(\text{wdraw}_m \mid d(t) = d_m) = \int_0^\infty V(\hat{\gamma}_w, \tau_w) f_I(x) \mathrm{d}x$$

where

$$\hat{\gamma}_w \triangleq 10 \log_{10} \left( \frac{\gamma_w d_m^\beta}{P_{tx}} (1 + x) \right)$$

and $f_I$ and $V$ are given by (39) and Proposition 4, respectively. $\blacksquare$

## VII. Applications

Using the above framework, we investigate the handover measurement in LTE and in particular the influence of key parameters on the system performance.

### A. System Scenarios

*1) Deployment Scenarios:* Parameters are summarized in Table II following 3GPP recommendations [4], [33] for two deployment scenarios of LTE networks, including urban and rural macrocellular networks. For each scenario, the network density $\lambda$ is set corresponding to hexagonal cellular layout of 3GPP standard, resulting in $\lambda = 2/(3\sqrt{3}R^2)$ BS/m$^2$.

The user's mobility is characterized in terms of his or her moving direction and velocity. The user is assumed to be moving away from the serving base station at velocity $v$. This scenario has been considered as the most critical circumstance [33]. The velocity is assumed constant and is set according to maximum speed authorized by regulations, typically 50 km/h in cities, and 130 km/h in highways.

TABLE III
SERVICES AND CONFIGURATION

| Parameter | | Assumption |
|---|---|---|
| Services | Min time to failure | $\tau_{\min} = 200$ ms |
| | Min level to failure | $\gamma_{\min} = -20$ to $-5$ dB |
| | Handover margin | $\delta_{\mathrm{HO}} = 2$ dB |
| | Req. target level | $\gamma_{\mathrm{req}} = \gamma_{\min} + \delta_{\mathrm{HO}}$ |
| Continual Measurement | Measurement period | $T_{\mathrm{meas}} = 200$ ms |
| | Triggering level | $\gamma_{\mathrm{t}} = +\infty$ |
| | Withdrawal level | $\gamma_{\mathrm{w}} = +\infty$ |
| Triggered Measurement | Measurement period | $T_{\mathrm{meas}} = 200$ ms |
| | Scan margin | $\delta_{\mathrm{scan}} = 20$ dB |
| | Triggering level | $\gamma_{\mathrm{t}} = \gamma_{\min} + \delta_{\mathrm{scan}}$ |
| | Withdrawal level | $\gamma_{\mathrm{w}} = \gamma_{\min} + \delta_{\mathrm{scan}}$ |
| | Min time to trigger | $\tau_{\mathrm{t}} = 200$ ms |
| | Min time to withdraw | $\tau_{\mathrm{w}} = 1024$ ms |

In lognormal shadowing, the square-exponential autocorrelation model [34], [35] is used

$$R_X(\tau) = \sigma_X^2 \exp\left(-\frac{1}{2}\left(\frac{v\tau}{d_c}\right)^2\right)$$

where $d_c$ is the decorrelation distance. This model satisfies conditions required in Section III. Its second spectral $\lambda_2$ is given by

$$\lambda_2 = -R_X''(\tau)|_{(\tau=0)} = \left(\frac{\sigma_X v}{d_c}\right)^2.$$

*2) Service Requirements:* The minimum allowable level $\gamma_{\min}$ and minimum duration to service failure $\tau_{\min}$ are set according to the condition of a *radio link failure* specified by LTE standard [4, Ch. 7.6]. When the downlink radio link quality estimated over the last 200-ms period becomes worse than a threshold $Q_{\mathrm{out}}$, Layer 1 of the UE shall send an *out-of-sync* indication to higher layers. Upon receiving $N_{310}$ consecutive *out-of-sync* indications from Layer 1, the UE will start timer $T_{310}$. Upon the expiry of this timer, the UE considers radio link failure to be detected. It can be consequently concluded that a radio link failure occurs if the signal quality of the serving cell is worse than $\gamma_{\min} = Q_{\mathrm{out}}$ during at least

$$\tau_{\min} = 200[\mathrm{ms}] \cdot N_{310} + T_{310}.$$

Following the parameters specified in [4, Ch. A.6], $N_{310} = 1$ and $T_{310} = 0$. This yields $\tau_{\min} = 200$ ms. The threshold $Q_{\mathrm{out}}$ is defined in [4, Ch. 7.6.1] as the level at which the downlink radio link cannot be reliably received and corresponds to 10% block error rate of physical downlink control channel (PDCCH). Note that $Q_{\mathrm{out}}$ is as small as $-10$ dB [33, Ch. A.2]. Moreover, various settings of $\gamma_{\min}$ are evaluated, as summarized in Table III.

The target cell's quality is required to be higher than the minimum tolerable level $\gamma_{\min}$ by a handover margin $\delta_{\mathrm{HO}}$ such that $\gamma_{\mathrm{req}} = \gamma_{\min} + \delta_{\mathrm{HO}}$.

*3) System Configuration:* The LTE standard assumes that UEs perform the handover measurement autonomously using 504 PCIs without NCL. The probability of finding a suitable handover target $\mathbb{P}(\mathrm{findtarget}_m(k))$ is thus given by Proposition 1 with retention probability $\rho_k = k/N_{\mathrm{CSID}}$, where $N_{\mathrm{CSID}} = 504$.

The conventional configuration of LTE standard specifies that a UE measures neighboring cells as soon as it enters connected-mode. This configuration is commonly referred to as *continual* handover measurement. This setting corresponds to triggering level and withdrawal level set to infinity, i.e., $\gamma_{\mathrm{t}} = +\infty$, and $\gamma_{\mathrm{w}} = +\infty$, so that $\pi_0(2) = 1$, and $\mathbb{P}\{\mathrm{wdraw}_m\} = 0$ according to (9). On the other hand

$$\mathbb{P}(\mathrm{trig}_m) = 1 - \mathbb{P}(\mathrm{fail}_m)$$

according to (42), for all $m$. This implies $\pi_m(1,1) = 0$ following (10). Note that in this case of continual measurement, the transition matrix reduces to

$$\mathbf{M}_m = \begin{pmatrix} 0 & \pi_m(1,2) & 0 & \pi_m(1,4) \\ 0 & \pi_m(2,2) & \pi_m(2,3) & \pi_m(2,4) \\ 0 & 0 & 1 & 0 \\ 0 & 0 & 0 & 1 \end{pmatrix}.$$

Beside the above conventional configuration, we also consider *triggered* handover measurement in which $\gamma_{\mathrm{t}}$ and $\gamma_{\mathrm{w}}$ are set to the same finite threshold given by

$$\gamma_{\mathrm{t}} = \gamma_{\mathrm{w}} = \gamma_{\min} + \delta_{\mathrm{scan}}$$

where $\delta_{\mathrm{scan}}$ is the scan margin. This setting helps in examining the influence of the system configurations and also showing the capability of the developed model. The parameters are summarized in Table III.

*B. Validation*

A computer simulation was built with the above urban macrocell scenario in order to check the accuracy of the models developed in Section VI. The interference field was generated according to a Poisson point process with intensity $\lambda$ in a 100-km$^2$ region, and the serving base station is located at the center of this region. The auto-correlated shadowing was generated as the output of an infinite impulse response filter with input Gaussian noise of standard deviation $\sigma_X$.

First, Fig. 5(a) verifies our analytical model against computer simulation for the tail distribution of the best signal quality $\bar{F}_{Y_k}$ of Proposition 1, which corresponds to common LTE setting described above. The agreement of the results by the proposed analytical expressions and simulations illustrates the accuracy of modeling the best signal quality $Y_k$ defined in (6) by the maximum of SINRs received from the thinning $S_k$ proposed in (24).

Fig. 5(b) checks the analytical framework based on level crossing analysis that was used to derive the probabilities of service failure, scanning triggering, and scanning withdrawal. Results show that both the analytical model and the simulation provide agreed results of the probability of service failure $\mathbb{P}(\mathrm{fail}_m)$ in both settings.

*C. Results*

With the accuracy provided by the proposed analytical framework, we investigate numerical results of the defined performance metrics for the following scenarios:

- Scenario 1: Rural with continual measurement;
- Scenario 2: Rural with triggered measurement;
- Scenario 3: Urban with continual measurement;
- Scenario 4: Urban with triggered measurement.

*1) Continual Measurement in Rural Macrocell:* Fig. 6 evaluates the performance of handover measurement in this scenario

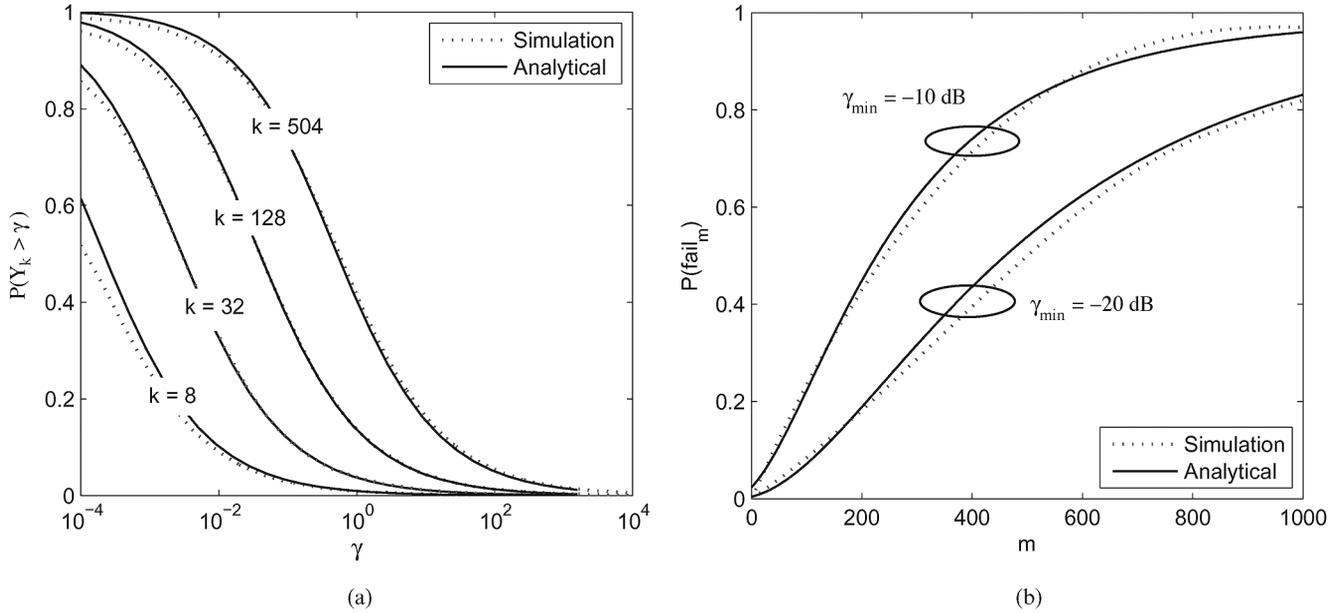

Fig. 5. Validation of analytical model. Network scenario is urban macrocell. (a) Tail distribution of $Y_k$. (b) Level crossing analysis.

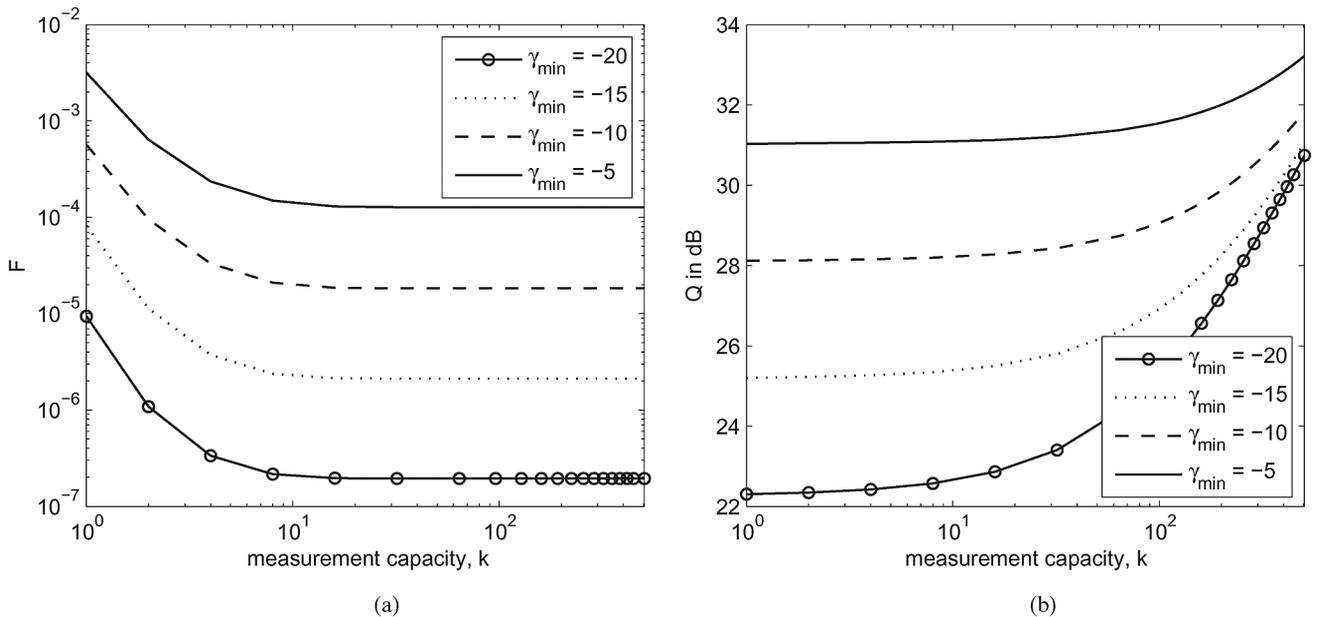

Fig. 6. Continual measurement in rural macrocell (Scenario 1). (a) Measurement failure probability $\mathcal{F}$. (b) Target cell quality $\mathcal{Q}$.

for different service requirements and measurement capacity. It shows that the handover measurement failure probability $\mathcal{F}$ and the expected quality of target cell $\mathcal{Q}$ are enhanced by increased measurement capacity $k$. This is by the fact that higher $k$ improves the distribution of the maximum SINR—cf. Fig. 5(a) or see [26] and [27] for analytical implication—resulting in higher probability that a UE will find a suitable target cell.

Fig. 6 also indicates the dependence of $\mathcal{F}$ and $\mathcal{Q}$ on the service requirement $\gamma_{\min}$. For more constrained services (i.e., higher $\gamma_{\min}$), the probability of service failure is higher, leading to higher probability of handover measurement failure as seen in Fig. 6(a). On the other hand, from Fig. 6(b), the quality of target cell $\mathcal{Q}$ is proportionally enhanced with higher required level of target cell quality $\gamma_{\mathrm{req}}$, which is set to $\gamma_{\min} + \delta_{\mathrm{HO}}$ with $\delta_{\mathrm{HO}} = 2$ dB in this scenario.

Note that the increasing rate of $\mathcal{Q}$ with high $\gamma_{\mathrm{req}} (= \gamma_{\min} + \delta_{\mathrm{HO}})$ is smaller than that of low $\gamma_{\mathrm{req}}$; see Fig. 6(b). Moreover, as shown in Fig. 6(a), $\mathcal{F}$ is below $10^{-2}$ and flattens out around $k = 8$ for all cases. Hence, for the continual handover measurement in a rural macrocell environment, supporting low measurement capacity could be enough for reliable handover measurement. This setting is also good for reducing mobile's power consumption. One can confirm that $k = 8$ recommended by 3GPP standard is indeed efficient. On the other hand, setting a relatively high $\gamma_{\mathrm{req}}$ could arrive a higher expected target cell quality $\mathcal{Q}$. However, the tradeoff is that it may lead to higher failure probability $\mathcal{F}$.

*2) Triggered Measurement in Rural Macrocell:* The continual measurement as seen above provides good performance in terms of low failure probability. However, it consumes ter-minal's battery by continuously processing cell measurement when in connected mode. An option for reducing this is to use triggered measurement.

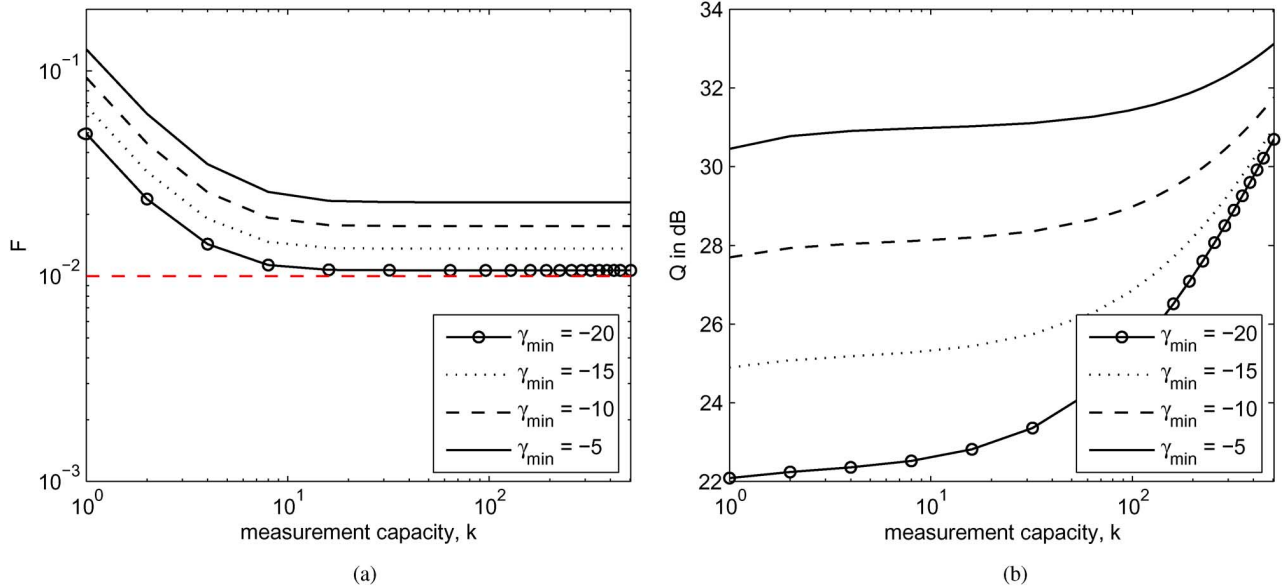

Fig. 7. Triggered measurement in rural macrocell (Scenario 2). (a) Measurement failure probability $\mathcal{F}$. (b) Target cell quality $\mathcal{Q}$.

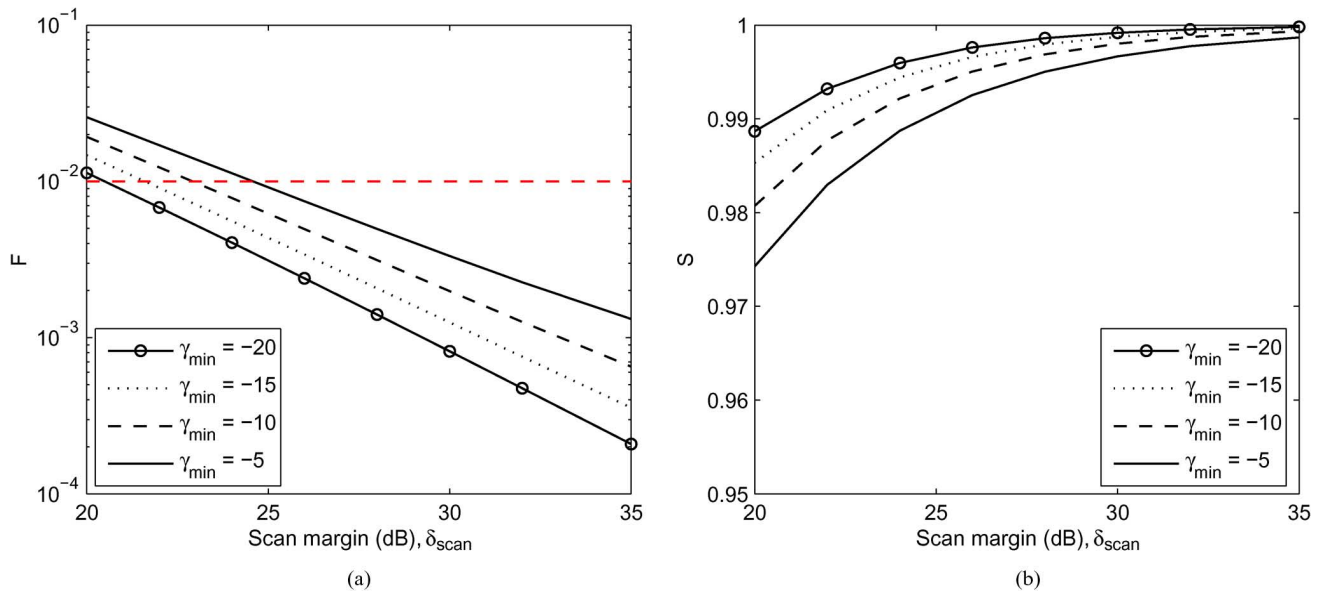

Fig. 8. Influence of scan margin $\delta_{\mathrm{scan}}$. Rural macrocell with measurement capacity $k = 8$. (a) Measurement failure probability. (b) Measurement success probability.

For the triggered measurement, Fig. 7 shows $\mathcal{F}$ and $\mathcal{Q}$ with respect to the measurement capacity while the service requirements are similar to those in continual measurement. Compared to the former case (see Fig. 6), the failure probability is clearly higher, however the target cell quality only has minor degradation.

Note that indeed the continual measurement is a triggered measurement with $\delta_{\mathrm{scan}} = +\infty$. Fig. 8 shows how triggered measurement could degrade the performance of handover measurement in terms of the scan margin $\delta_{\mathrm{scan}}$. The higher $\delta_{\mathrm{scan}}$ is, the more room the mobile can have to find a target cell before a service failure, resulting in better handover measurement performance in terms of lower failure probability and higher success probability. Moreover, comparing Fig. 6(a) at $k = 8$ and Fig. 8(a) at $\delta_{\mathrm{scan}} = 20$, we can see that $\mathcal{F}$ increases by a factor of more than $10^2$ when moving from continual measurement to triggered measurement. This explains significant degradation of $\mathcal{F}$ observed in Fig. 7(a) compared to Fig. 6(a).

On the other hand, Fig. 8(b) shows that the success probability $\mathcal{S}$ is already near to 1. The influence on $\mathcal{S}$ when changing $\delta_{\mathrm{scan}}$ from $+\infty$ (i.e., continual measurement) to 20 dB is less significant. This results in relatively small degradation of $\mathcal{Q}$.

The counterpart is that the mobile performs handover measurement more frequently when setting higher scan margin $\delta_{\mathrm{scan}}$, leading to more power consumption. One question of interest thus is the optimal setting of $\delta_{\mathrm{scan}}$ to get a good tradeoff between the handover measurement performance and the power consumption. As far as the information on the power consumption due to handover measurement is unavailable, it can be efficient to set $\delta_{\text{scan}}$ to the minimum value for which the failure probability is below a target level, e.g., $10^{-2}$.

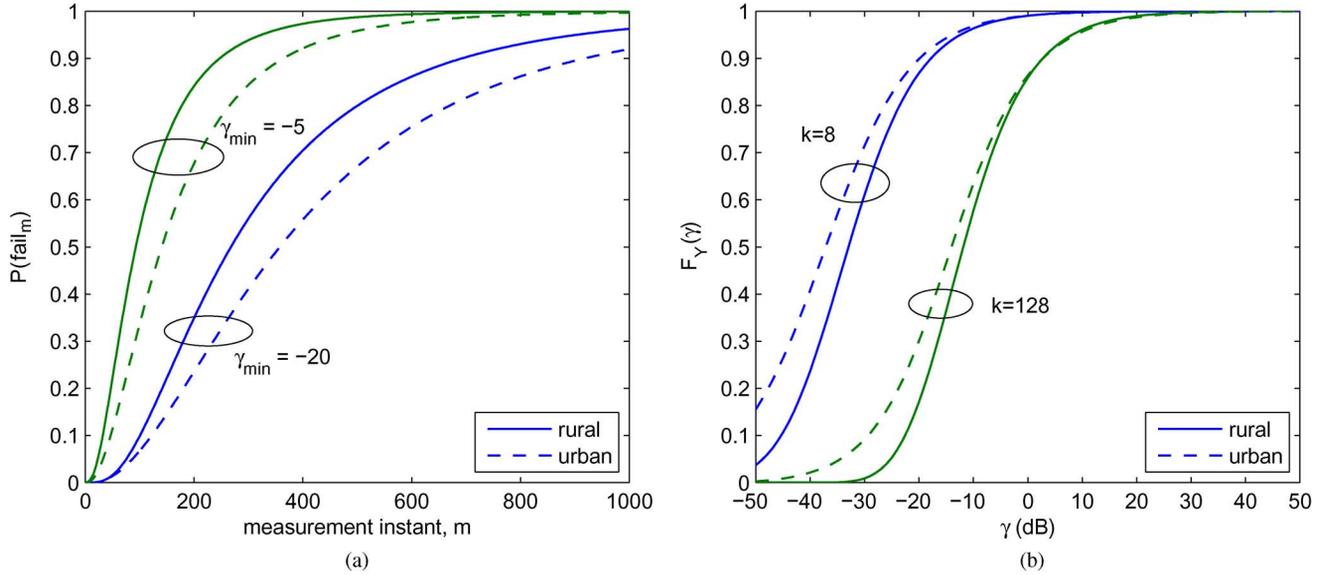

Fig. 9.   Urban macrocell versus rural macrocell. (a) Service failure probability. (b) Distribution of maximum SINR.

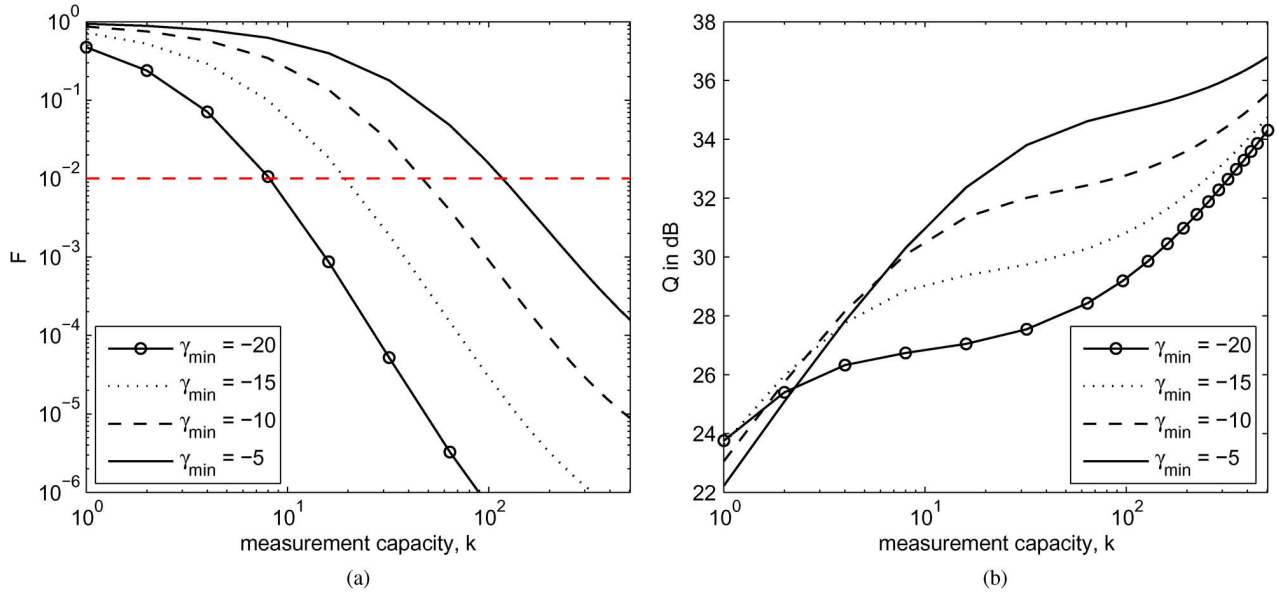

Fig. 10.   Continual measurement in urban macrocell (Scenario 3). (a) Measurement failure probability. (b) Target cell quality.

*3) Continual Measurement in Urban Macrocell:* We now assess the influence of the network environment and user's mobility on the handover measurement by considering urban macrocell network in which the user's velocity is lower than that in the rural macrocell case, cf. Table II.

As seen from the analytical results, the failure probability $\mathcal{F}$ depends on various parameters including the distribution of the maximum SINR and the service failure probability. The latter in turn depends on the user's mobility. Fig. 9(a) shows that the service failure probability is higher in the rural case than in the urban one. However, as the distribution of the maximum SINR in the urban macrocell is worse, cf. Fig. 9(b), it results in higher probability $\mathcal{F}$ compared to that in the rural macrocell case, cf.

Figs. 10(a) and 6(a). For the same target level of $\mathcal{F}$ at $10^{-2}$ as above, measurement capacity $k = 8$ could be only sufficient for robust services (i.e., low $\gamma_{\text{min}}$); see Fig. 10(a). To support more constrained services in urban macrocell environment, the mobile terminal should be able to measure as many as 100 cells per measurement period of 200 ms.

In Fig. 10(b), we see that there is a crossing point between the curves when the measurement capacity goes around $k = 2$. Given the distribution of the maximum SINR, we know that the target cell quality $\mathcal{Q}$ is proportional to both the success probability $\mathcal{S}$ and the resulting quality of the target cell $\mathbb{E}\{Y_k | Y_k \geq \gamma_{\text{req}}\}$, cf. V-F. With $\gamma_{\text{req}} = \gamma_{\text{min}} + \delta_{\text{HO}}$ where $\delta_{\text{HO}}$ is 2 dB in the current evaluation, $\mathbb{E}\{Y_k | Y_k \geq \gamma_{\text{req}}\}$ is proportional to $\gamma_{\text{min}}$. However, $\mathcal{S}$ is inversely proportional to $\gamma_{\text{min}}$. Thus, the characteristics of $\mathcal{Q}$ when increasing $\gamma_{\text{min}}$ depend on how effectively $\mathbb{E}\{Y_k | Y_k \geq \gamma_{\text{req}}\}$ compensates the degradation in $\mathcal{S}$. Fig. 10(b) indicates that for small measurement capacity $k$, it is better to maintain low $\gamma_{\mathrm{req}}$, either by privileging robust services or by setting low handover margin $\delta_{\mathrm{HO}}$. However, $\gamma_{\mathrm{min}}$ is an intrinsic requirement of the service; the remaining option is to fine-tune the handover margin to gain an optimal operation.

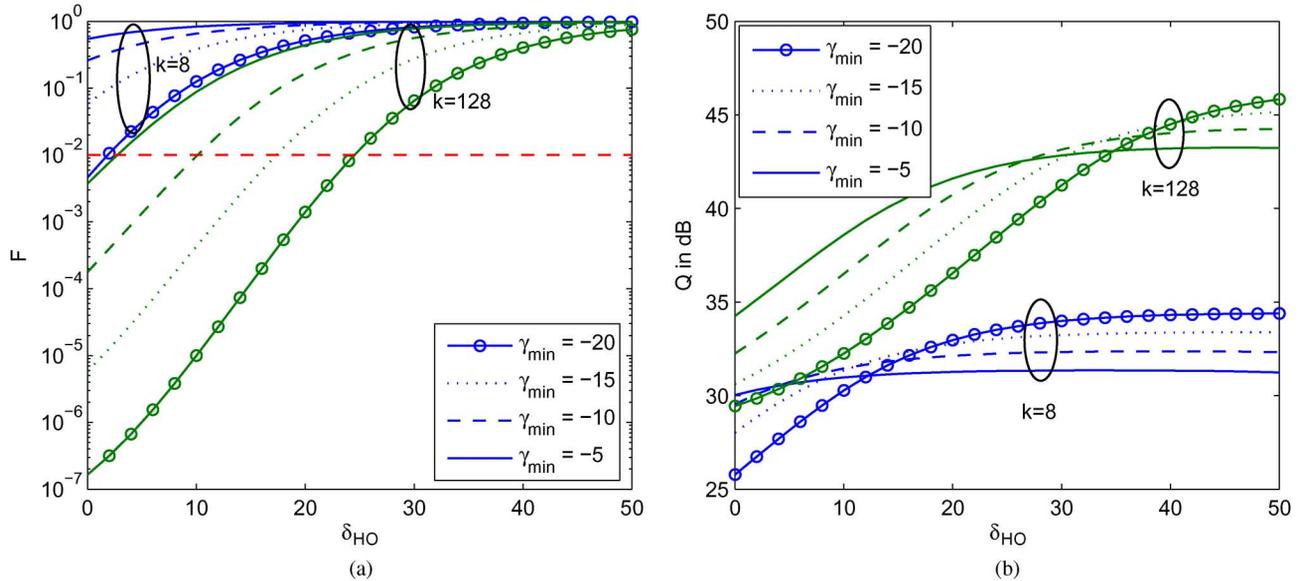

Fig. 11. Influence of $\delta_{\mathrm{HO}}$: continual measurement in urban macrocell. (a) Measurement failure probability. (b) Target cell quality.

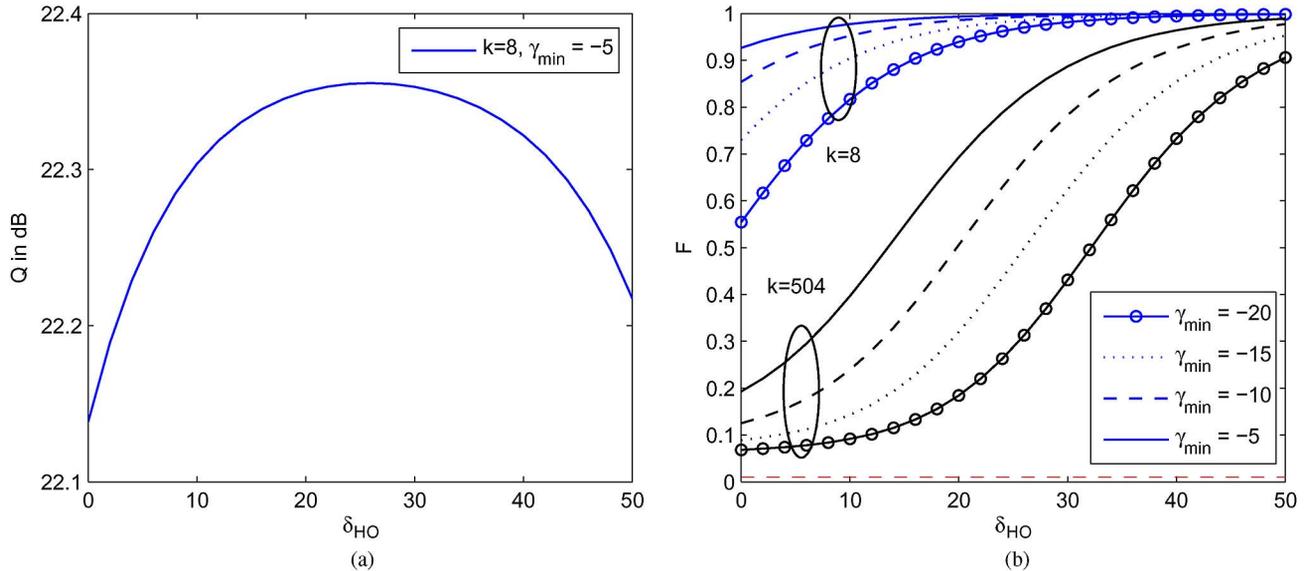

Fig. 12. Influence of handover margin $\delta_{\mathrm{HO}}$: evaluation with triggered measurement of $\delta_{\mathrm{scan}} = 20$ dB in urban macrocell. (a) Optimal $\delta_{\mathrm{HO}}$. (b) Referring failure probability $\mathcal{F}$.

By the above consideration, for a fixed $\gamma_{\mathrm{min}}$, $\mathcal{Q}$ should be concave with respect to $\delta_{\mathrm{HO}}$. Fig. 11 assesses the influence of the handover margin $\delta_{\mathrm{HO}}$ in the case of continual measurement. First of all, for all services (i.e., $\gamma_{\mathrm{min}}$), increasing $\delta_{\mathrm{HO}}$ results in higher failure probability $\mathcal{F}$, cf. Fig. 11(a). This is by the fact that the probability of finding a target cell $\mathbb{P}(Y_k \geq \gamma_{\mathrm{min}} + \delta_{\mathrm{HO}})$ decreases as $\delta_{\mathrm{HO}}$ increases. With the assessed range of $\delta_{\mathrm{HO}}$, the target cell quality $\mathcal{Q}$ of all services is proportional to $\delta_{\mathrm{HO}}$. cf. Fig. 11(b). However, the more demanding the service is, the lower the gain of $\mathcal{Q}$ will be when increasing $\delta_{\mathrm{HO}}$.

*4) Triggered Measurement in Urban Macrocell:* We continue the assessment of the influence of handover margin $\delta_{\mathrm{HO}}$. In the continual measurement case considered above, the maximum of $\mathcal{Q}$ is still outside the evaluated range of $\delta_{\mathrm{HO}}$. In the case of triggered measurement, we can see the concave behavior of $\mathcal{Q}$ with respect to $\delta_{\mathrm{HO}}$ in Fig. 12(a). Ideally, the handover margin should be set to the value at the maximum of $\mathcal{Q}$. However, we see from Fig. 12(b) that even at the maximum measurement capacity $k = 504$, the failure probability $\mathcal{F}$ is still too high for all services and all the possible settings of $\delta_{\mathrm{HO}}$. As a consequence, one could recommend to use continual handover measurement in an urban macrocell network to secure reliable mobility supports.

## VIII. CONCLUSION

In a mobile cellular network, handover measurement provides the mobile station with necessary controls to find a suitable handover target to which it can be switched when the current cell's signal deteriorates. It directly decides the quality of the handover target and has strong impact on the ongoing services. Through this paper, we developed a unified framework to analyze this function for multicell systems. Essentially, a handover measurement procedure is characterized by four key probabilistic events including: 1) suitable handover target found; 2) service failure; 3) measurement triggering; and 4) measurement withdrawal. Built on these probabilistic events, we represent its temporal evolution for analytical performance evaluation. We derived closed-form results for the transition probabilities taking into account user's dynamics and the system control. The developed framework unifies the influences of various parameters. Most importantly, it can be used to optimize system configuration under different scenarios.

We showed applications of the model to the handover measurement in 3GPP LTE systems, when the neighbor cell list is not available to UE. Results showed the following.

- With continual handover measurement, the current standard's requirement on UE's measurement capability of measuring eight cells per 200 ms is sufficient to guarantee good measurement performance in a rural macrocell network. Configuring lower measurement capability is even possible to reduce terminal's battery consumption.
- Using triggered handover measurement in a rural macrocell network is also possible for further reduction of the terminal's battery consumption. It is necessary to configure the scan margin efficiently to keep a good tradeoff with the handover measurement performance.
- However, in a network like urban macrocell where the signal suffers from more interference, it is necessary to use continual measurement since triggered measurement can result in high probability of measurement failure. Moreover, the UE's measurement capability should be as high as 100 cells per 200 ms to support more demanding services.
- Handover margin offers another degree of freedom to optimize the handover measurement. It should be set so as to maximize the target cell quality while keeping low level of the handover measurement failure.

## Appendix A
### Asymptotic Properties of Excursions

*Lemma 10 ([36, Theorem 10.4.2]):* With the process $X(t)$ as described above, if $R_X(\tau)$ satisfies (35) and (37), then excursions of $X(t)$ above $\gamma$ behave asymptotically as

$$X(t) \sim \gamma + \xi t - \gamma \frac{\lambda_2 t^2}{2}, \qquad \text{as } \gamma \to +\infty \qquad (46)$$

where $\xi$ is a Rayleigh random variable of parameter $\sigma_\xi = \sqrt{\lambda_2}$.

*Lemma 11 ([32]):* With the process $X(t)$ described above, let $T$ be the time of an up-excursion of $X(t)$ above a level $\gamma$. If $R_X(\tau)$ satisfies (36) and (37), then $T$ asymptotically follows an exponential distribution of rate $\mu = \mathbb{E}U_\gamma$ as $\gamma \to -\infty$, i.e.,

$$\mathbb{P}(T \leq \tau) = 1 - e^{-\mu\tau}, \qquad \text{as } \gamma \to -\infty.$$

## Appendix B
### Proof of Proposition 4

Let $T$ be the time interval of an excursion of $X(t)$ above $\gamma$, and $\mathcal{E}_\gamma(\mathcal{I})$ be the number of excursions of $X(t)$ above $\gamma$ during interval $\mathcal{I}$. We have

$$V(\gamma,\tau) = \lim_{\mathcal{I}\to\infty} \frac{\mathcal{E}_\gamma(\mathcal{I})\mathbb{P}(T \geq \tau)\mathbb{E}(T|T \geq \tau)}{\mathcal{I}} \qquad (47)$$

where $\mathbb{E}(T|T \geq \tau)$ is the average interval of an excursion above $\gamma$ given that the excursion lasts for at least $\tau$, and the numerator on the right-hand side is nothing but the total time that is accumulated each time $X(t)$ stays above $\gamma$ during at least $\tau$.

Note that under the conditions considered for $R_X(\tau)$, the Gaussian process $X(t)$ is equivalent to some process $\eta(t)$ that has continuous sample paths (and moreover derivative), with probability one [23, Eq. (9.5.4)]. Hence, the number of excursions above $\gamma$ is equal to the number of up-crossings of $\gamma$

$$\lim_{\mathcal{I}\to\infty} \frac{\mathcal{E}_\gamma(\mathcal{I})}{\mathcal{I}} = \mathbb{E}U_\gamma.$$

Introducing this back to (47) yields

$$V(\gamma,\tau) = \mathbb{E}U_\gamma \times \mathbb{P}(T \geq \tau) \times \mathbb{E}(T|T \geq \tau). \qquad (48)$$

### A. For $\gamma \to +\infty$

The asymptotic trajectory of an excursion of $X(t)$ above $\gamma$ can be described by Lemma 10. Using (46) and solving for equation $X(t) = \gamma$, we obtain two solutions, say $t_1$ and $t_2$, and have $T$ given by $|t_2 - t_1|$, which is equal to

$$T = \frac{2}{\gamma\lambda_2}\xi.$$

As a result, the asymptotic mean of $T$ is

$$\mathbb{E}T_{\text{asymp}} = \frac{2}{\gamma\lambda_2}\mathbb{E}\xi = \frac{2}{\gamma\lambda_2}\sqrt{\frac{\pi}{2}}\sigma_\xi.$$

However, the exact mean of $T$ is given as

$$\mathbb{E}T = \frac{\mathbb{P}(X \geq \gamma)}{\mathbb{E}U_\gamma} = \frac{Q\left(\frac{\gamma}{\sigma_X}\right)}{\mathbb{E}U_\gamma}$$

where $Q(x) = (1/\sqrt{2\pi})\int_x^\infty \exp(-(x^2/2))\mathrm{d}x$. The first-order estimation of $T$ should imply that $\mathbb{E}T_{\text{asymp}} = \mathbb{E}T$, leading to

$$\sigma_\xi = \frac{\gamma\lambda_2}{\sqrt{2\pi}}\frac{Q\left(\frac{\gamma}{\sigma_X}\right)}{\mathbb{E}U_\gamma}.$$

Note that for $\gamma \to +\infty$, using approximation $Q(x) \simeq \exp(-x^2/2)/(x\sqrt{2\pi})$ as $x \to +\infty$, we can easily find back $\sigma_\xi = \sqrt{\lambda_2}$ of Lemma 10.

Thus, the tail distribution of $T$ is given by that of $\xi$ as follows:

$$\mathbb{P}(T > t) = \mathbb{P}\left(\xi > \frac{\gamma\lambda_2}{2}t\right) = \exp\left(-A_\gamma t^2\right) \qquad (49)$$

where $A_\gamma := (\gamma\lambda_2/2)^2/(2\sigma_\xi^2)$.

Noting that $\mathbb{P}(T \geq t) = \mathbb{P}(T > t)$ as $\mathbb{P}(T > t)$ is continuous, we have

$$\mathbb{E}\{T|T \geq \tau\} = \int_0^{+\infty} \mathbb{P}(T > t|T \geq \tau)\mathrm{d}t$$

$$= \tau + \int_\tau^{+\infty} \frac{\mathbb{P}(T > t)}{\mathbb{P}(T \geq \tau)}\mathrm{d}t$$

$$= \tau + \sqrt{\frac{\pi}{A_\gamma}}\frac{Q(\sqrt{2A_\gamma}\tau)}{\mathbb{P}(T > \tau)}. \qquad (50)$$

Substituting (49) and (50) into (48), we get $V$ for $\gamma \to +\infty$.

### B. For $\gamma \to -\infty$

The asymptotic distribution of $T$ can be given by Lemma 11. Hence, by its exponential distribution, we can easily get

$$\mathbb{E}\{T|T \geq \tau\} = \frac{\tau + 1}{\mu}.$$

Also

$$V = \mathbb{E}U_\gamma \cdot \exp(-\mu\tau) \cdot \left(\frac{\tau + 1}{\mu}\right).$$

Like the case of positive $\gamma$, the asymptotic mean of $T$ is

$$\mathbb{E}T_{\text{asymp}} = \frac{1}{\mu}.$$

The first-order estimation of $T$ should imply that

$$\frac{1}{\mu} = \frac{Q\left(\frac{\gamma}{\sigma_X}\right)}{\mathbb{E}U_\gamma}.$$

Here, note that we find back $\mu \to \mathbb{E}U_\gamma$ as $\gamma \to -\infty$ as given by Lemma 11. This completes the proof. ∎

### ACKNOWLEDGMENT

A part of this work was realized at the research project on Network Theory and Communications (TREC) of INRIA-Ecole Normale Supérieure. The authors would like to thank Prof. F. Baccelli for his valuable discussions and supports.

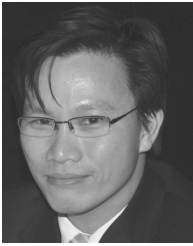

**Van Minh Nguyen** (M'11) received the M.Eng. degree in telecommunications and M.Sc. degree in communication networks from Telecom Bretagne, Brest, France, in 2007, the Certificate in Management from ENPC MBA Paris, Paris, France, in 2009, and the Ph.D. degree in mobile communications theory from Telecom ParisTech, Paris, France, in 2011.

Since 2011, he has been with the R&D Department, Sequans Communications, Paris, France, as a Research Scientist, and since 2012 has held in parallel a Senior DSP Engineer position. Prior to that, in 2007 he joined the Networks and Networking Department, Alcatel-Lucent Bell Labs, Nozay, France, where he was a Member of Technical Staff. In 2010, he was a Research Engineer in the research project on Network Theory and Communications (TREC) of INRIA-Ecole Normale Supérieure. His current research interests include mobile communications and networking, radio resource management, signal processing, MIMO, and wireless communications theory.

Dr. Nguyen has served as TPC member of international conferences including IEEE ICC, GLOBECOM, GreenCom, PIMRC, and ICCVE, and was the Session Chair of PIMRC. He received the Award of the French Embassy in Vietnam for the First Prize of the PFIEV Contest, the Award of Nortel Vietnam, and the Award of Alcatel CIT France for the best engineer of the year.

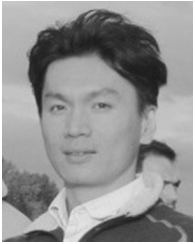

**Chung Shue Chen** (S'02–M'05) received the B.Eng., M.Phil., and Ph.D. degrees in information engineering from the Chinese University of Hong Kong, Hong Kong, in 1999, 2001, and 2005, respectively.

He joined Alcatel-Lucent Bell Labs, Nozay, France, in 2011, where he is a Member of Technical Staff. He also holds a position of permanent member with the Laboratory of Information, Networking and Communication Sciences (LINCS), Paris, France. Prior to that, he worked at INRIA, Paris, France, in the French National Institute for Research in Computer Science and Control, in the research group on Network Theory and Communications. He was an ERCIM Fellow with the Norwegian University of Science and Technology (NTNU), Trondheim, Norway, and the National Center for Mathematics and Computer Science (CWI), Amsterdam, The Netherlands. He was an Assistant Professor with the Chinese University of Hong Kong. His research interests include wireless communications and networking, radio resource management, self-organizing networks, and green communication systems.

Dr. Chen has served as TPC in international conferences including IEEE ICC, GLOBECOM, WCNC, VTC, CCNC, INFOCOM workshop on Cognitive and Cooperative Networks, WiOpt, etc. He has been an Editor of the *European Transactions on Telecommunications* since 2011. He received the ERCIM "Alain Bensoussan" Fellowship from the European Research Consortium for Informatics and Mathematics and the Sir Edward Youde Memorial Fellowship from Hong Kong.

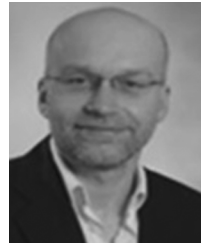

**Laurent Thomas** received the M.Eng. degree in telecommunications from the Institute Galilée of University of Paris Nord, Villetaneuse, France, in 1989.

In 2007, he joined Alcatel-Lucent Bell Labs, Nozay, France, where he has been Head of the Wireless IP and Mobile Networks Department. Prior to that, he worked in the R&D of cellular networks starting before GSM apparition with CT2 (Cordless Telephony 2 Standard) and Telepoint. He was head of the project on network technical optimization tools and procedures of SFR, where he designed and implemented one of the first statistical optimization systems of GSM networks with Metrica company. In 1999, he joined the R&D Department of Alcatel-Lucent for wireless systems operation and optimization tools. His research interests include network capacity and general-purpose processors for signal processing.